\documentclass[runningheads]{llncs}
\usepackage{amsmath,amssymb,mathtools,bm,booktabs}
\usepackage[linesnumbered,ruled,vlined]{algorithm2e}
\SetKw{Break}{break}

\usepackage{todonotes}
\definecolor{myred}{cmyk}{0,0.7,0.78,0.25}
\definecolor{myblue}{cmyk}{0.79,0.48,0,0.62}
\definecolor{myyellow}{cmyk}{0,0.02,0.23,0.01}

\mathtoolsset{showonlyrefs}
\usepackage[numbers, sort]{natbib}
\usepackage{tikz}
\usetikzlibrary{positioning,shapes.misc,calc}
\newcommand*\marked[1]{\tikz[baseline=(char.base)]{\node[shape=rectangle,draw=yellow,fill=myyellow,inner sep=2pt] (char) {#1};}}

\usepackage{comment}
\usepackage{etoolbox}
\usepackage{version}

\usepackage{fullpage}

\newcommand{\ot}{\leftarrow}
\newcommand{\cD}{\mathcal{D}}
\newcommand{\OPT}{\mathrm{OPT}}
\newcommand{\Opt}{\mathrm{Opt}}
\newcommand{\ALG}{\mathrm{ALG}}

\newcommand{\Random}{\textsc{Random}}
\renewcommand{\epsilon}{\varepsilon}
\newcommand{\ind}{\mathrm{ind}}
\renewcommand{\mid}{\,:\,}

\DeclareMathOperator*{\argmax}{arg\,max}

\DeclareMathOperator{\type}{type}

\title{Online Max-min Fair Allocation}
\author{Yasushi Kawase\inst{1} \and Hanna Sumita\inst{2}}
\authorrunning{Y. Kawase and H. Sumita}
\institute{University of Tokyo, Tokyo, Japan \email{kawase@mist.i.u-tokyo.ac.jp}\and 
Tokyo Institute of Technology, Tokyo, Japan \email{sumita@c.titech.ac.jp}}

\begin{document}
\maketitle

\begin{abstract}
  We study an online version of the max-min fair allocation problem for indivisible items.
  In this problem, items arrive one by one, and each item must be allocated irrevocably on arrival to one of $n$ agents, who have additive valuations for the items.
  Our goal is to make the least happy agent as happy as possible.
  In research on the topic of online allocation, this is a fundamental and natural problem.
  Our main result is to reveal the asymptotic competitive ratios of the problem for both the adversarial and i.i.d.~input models.
  We design a polynomial-time deterministic algorithm that is asymptotically $1/n$-competitive for the adversarial model, and we show that this guarantee is optimal.
  To this end, we present a randomized algorithm with the same competitive ratio first and then derandomize it. A natural derandomization fails to achieve the competitive ratio of $1/n$. We instead build the algorithm by introducing a novel technique.
  When the items are drawn from an unknown identical and independent distribution, we construct a simple polynomial-time deterministic algorithm that outputs a nearly optimal allocation.
  We analyze the strict competitive ratio and show almost tight bounds for the solution.
  We further mention some implications of our results on variants of the problem.
\end{abstract}

\section{Introduction}
% specific situation
In this paper, we study the problem of allocating indivisible items so that the minimum happiness among agents is maximized.
Let us consider a toy instance.
Suppose that Alice and Bob are trying to share bite-sized snacks that arrive sequentially.
As soon as each snack arrives, one of them will receive and eat it.
If each snack is picked by the one who values it more than the other, the outcome will become an imbalanced one (Table~\ref{tbl:prefer}).
In contrast, if they pick the items alternately, the outcome will become an inefficient one (Table~\ref{tbl:alternatively}).
The question then arises as to what kind of rule would satisfy fairness and efficiency simultaneously, and moreover, what would be the best possible rule.

\begin{table}[htb]
  \begin{minipage}{.45\textwidth}
    \centering
    \caption{Outcome when the snack is picked by the one who values it more}\label{tbl:prefer}
    \scalebox{1}{
      \tabcolsep=1mm
      \begin{tabular}{c|ccccccc}
        \toprule
                      & 1              & 2              & 3              & 4              & 5              & 6              & $\cdots$ \\\midrule
        Alice's value & \marked{$0.7$} & \marked{$1.0$} & \marked{$0.8$} & \marked{$0.9$} & \marked{$0.7$} & \marked{$0.8$} & $\cdots$ \\
        Bob's value   & $0.5$          & $0.1$          & $0.7$          & $0.2$          & $0.6$          & $0.0$          & $\cdots$ \\
        \bottomrule
      \end{tabular}
    }
  \end{minipage}
  \qquad
  \begin{minipage}{.45\textwidth}
    \centering
    \caption{Outcome when the snack is picked alternately}\label{tbl:alternatively}
    \scalebox{1}{
      \tabcolsep=1mm
      \begin{tabular}{c|ccccccc}
        \toprule
                      & 1              & 2              & 3              & 4              & 5              & 6              & $\cdots$ \\\midrule
        Alice's value & \marked{$0.7$} & $1.0$          & \marked{$0.8$} & $0.9$          & \marked{$0.7$} & $0.8$          & $\cdots$ \\
        Bob's value   & $0.5$          & \marked{$0.1$} & $0.7$          & \marked{$0.2$} & $0.6$          & \marked{$0.0$} & $\cdots$ \\
        \bottomrule
      \end{tabular}
    }
  \end{minipage}
\end{table}

% history
The fair allocation of resources or items to agents has been a central problem in economic theory for several decades.
In classical fair allocation problems, we are given all the items in advance.
Recently, the problem of allocating items in an online fashion has been studied in the areas of combinatorial optimization, algorithmic game theory, and artificial intelligence.
In online problems, indivisible items arrive one by one, and they need to be allocated immediately and irrevocably to agents.
The study of online fair allocation is motivated by its wide range of applications such as the allocation of donor organs to patients, donated food to charities, electric vehicles to charging stations; we refer the reader to the survey~\cite{Aleksandrov2020} for details.

% model
Throughout the paper, we denote the sets of agents and indivisible items by $N=\{1,2,\dots,n\}$ and $M=\{e_1,e_2,\dots,e_m\}$, respectively.
We use the symbol $[n]$ to denote $\{1,2,\dots,n\}$.
Each agent has a valuation function $v_i\colon M\to[0,1]$ that assigns a value to each item.
For simplicity, unless otherwise stated, we assume that the value of each item is normalized to $[0,1]$.
%For simplicity, we basically assume that the value of each item is normalized to $[0,1]$. We will also discuss algorithms without this assumption later.
We assume that each agent has an additive preference over the items, and we write $v_i(X)\coloneqq\sum_{e\in X}v_i(e)$ to denote the utility of agent $i$ when $i$ obtains $X\subseteq M$.
For an item $e\in M$, we call $(v_1(e),\dots,v_n(e))$ the \emph{value vector} of $e$.
An allocation $A=(A_1,\dots,A_n)$ is a partition of $M$ (i.e., $\bigcup_{i}A_i=M$ and $A_i\cap A_j=\emptyset$ for any distinct $i,j\in N$).
For $j\in[m]$, we denote $M^{(j)}=\{e_1,\dots,e_j\}$ and $A^{(j)}=(A_1\cap M^{(j)},\dots,A_n\cap M^{(j)})$.

% our target
Our goal is to find an allocation $A$ that maximizes the minimum utility among the agents $\min_{i\in N}v_i(A_i)$.
The value $\min_{i\in N}v_i(A_i)$ is called the \emph{egalitarian social welfare} of allocation $A$.
The problem of maximizing the egalitarian social welfare when items arrive one by one is called the \emph{online max-min fair allocation problem}.
Here, we assume that the number of items is unknown in advance.
The max-min fairness (that is, the egalitarian social welfare is maximized) is one of the most commonly used notions for measuring fairness and efficiency, and it has been studied extensively in the area of fair allocation~\cite{Golovin2005,AS2010,Bouveret2016,Li2019,KS2020}.
Thus, our problem naturally models the above applications using the notion of max-min fairness.
We measure the performance of online algorithms using the \emph{competitive ratio}, which is the ratio of the egalitarian social welfare obtained by an online algorithm to that of the offline optimal value.
Furthermore, we consider two types of competitive ratio: \emph{strict} and \emph{asymptotic}.
In the strict setting, we consider the worst-case ratio for every possible input sequence,
whereas in the asymptotic setting, we consider the worst-case ratio for input sequences with sufficiently large optimal values.
Section~\ref{sec:preliminary-cr} presents the formal definitions for these terms.
Note that the asymptotic competitive ratio represents an intrinsic performance ratio that does not depend on initial behavior.
We consider two arrival models: \emph{adversarial}, in which the items are chosen arbitrarily, and \emph{independent and identically distributed (i.i.d.)}, in which the value vectors of the items are drawn independently from an unknown/known distribution.
Note that a value vector can be a continuous random variable in the i.i.d.~arrival model.

\subsection{Related work}
% Load balancing
A class of the online max-min fair allocation problem with identical agents (i.e., $v_1=\cdots=v_n$) has also been studied as the \emph{online machine covering problem} in the context of scheduling~\cite{Deuermeyer1982,Woeginger1997,Azar1998,Tan2013,Galvez2015,Galvez2020}.
Here, an agent's utility corresponds to a machine load.
The problem of maximizing the minimum machine load was initially motivated by modeling the sequencing of maintenance actions for modular gas turbine aircraft engines~\cite{Deuermeyer1982}.
For this case, it is known that any online deterministic algorithm has a strict competitive ratio of at most $1/n$ and that the greedy algorithm is strictly $1/n$-competitive~\cite{Woeginger1997}.
Besides, there exists a strictly $\Omega(\frac{1}{\sqrt{n}\log n})$-competitive randomized algorithm, which is a best possible algorithm up to logarithmic factors~\cite{Azar1998}.

% vanish envy and other properties
In addition to the online max-min fair allocation problem, online fair allocation problems with other fairness and efficiency notions have been studied~\cite{Aleksandrov2015,Aleksandrov2017a,Aleksandrov2017b,Aleksandrov2017c,Mattei2017,Benade2018,Aleksandrov2019,Aleksandrov2020,Bogomolnaia2021}.
For example, Benade et al.~\cite{Benade2018} focused on an online problem of allocating all the indivisible items to minimize the maximum envy.
They designed a deterministic online algorithm such that the maximum envy is sublinear with respect to the number of items; the algorithm outputs an allocation $A$ such that $v_i(A_i)\ge v_i(A_j)-O(\sqrt{m\log m})$ for any $i,j\in N$.
Unlike our setting, they assumed that the number of items is known in advance.
Their algorithm is based on a random allocation, where each item is allocated to an agent chosen uniformly at random.
In \cite{Benade2018}, the authors first prove that the maximum envy in the allocation obtained by the random allocation algorithm is sublinear.
Then, they derandomized the algorithm by using a potential function that pessimistically estimates the future allocation.
For more models of online fair allocation, see \cite{Aleksandrov2020} for a comprehensive survey.

% Offline version
The offline version of the max-min allocation problem has also been studied under the name of the \emph{Santa Clause problem}~\cite{BezakovaD2005,Golovin2005,Feige2008,Chakrabarty2009,Haeupler2011}.
The problem is NP-hard even to approximate within a factor of better than $1/2$~\cite{Lenstra1990}.
Bansal and Sviridenko~\cite{BansalS2006} proposed an $\Omega(\log\log\log n/\log\log n)$-approximation algorithm for the restricted case when $v_i(e)\in\{0,v(e)\}$ for all $i\in N$ and $e\in M$.
Asadpour and Saberi~\cite{AS2010} provided the first polynomial-time approximation algorithm for the general problem, which was improved by Haeupler et al.~\cite{Haeupler2011}.

\subsection{Our results}
Although the online max-min fair allocation problem is a fundamental problem, almost nothing is known about the competitive analysis for nonidentical agents to the best of our knowledge.

Our main results show the asymptotic competitive ratios of optimal online algorithms for the adversarial and i.i.d.~arrival models.
In addition, we roughly identified the strict competitive ratios of optimal online algorithms, which are much smaller than those of the asymptotic ones.
We summarize our results in Table~\ref{tbl:summary}.
%for the online max-min fair allocation problem.

\begin{table}[ht]
  \centering
  \caption{Summary of our results for the online max-min fair allocation problem.
    %The results marked with a star (*) is tight up to a logarithmic factor.
    All values in the table represent both upper and lower bounds of the competitive ratios of optimal online algorithms, where $\tilde{\Theta}$ is a variant of big-Theta notation ignoring logarithmic factors.}\label{tbl:summary}
  \scalebox{1}{
    \begin{tabular}{r|llll}
      \toprule
              & Adversarial (det.)                                                               & Adversarial (rand.)                                                                           & Unknown i.i.d.                                                                                        & Known i.i.d.                                                                                          \\\midrule
      Strict  & $0$ {\scriptsize(Thm.~\ref{thm:det_upper})}                                      & $1/n^{\Theta(n)}$ {\scriptsize(Thms.~\ref{thm:rand_strict_lower}, \ref{thm:oblivious_upper})} & $1/e^{\tilde{\Theta}(n)}$ {\scriptsize(Thms.~\ref{thm:rand_strict_lower}, \ref{thm:upper_known_iid})} & $1/e^{\tilde{\Theta}(n)}$ {\scriptsize(Thms.~\ref{thm:rand_strict_lower}, \ref{thm:upper_known_iid})} \\
      Asympt. & $1/n$ {\scriptsize(Thms.~\ref{thm:det_lower}, \ref{thm:upper_asymp_adverarial})} & $1/n$ {\scriptsize(Thms.~\ref{thm:det_lower}, \ref{thm:upper_asymp_adverarial})}              & $1$ {\scriptsize(Thm.~\ref{thm:iid_lower})}                                                           & $1$ {\scriptsize(Thm.~\ref{thm:iid_lower})}                                                           \\\bottomrule
    \end{tabular}
  }
\end{table}
\subsubsection{Adversarial arrival model}
% asymptotic
A main result for the adversarial arrival model is a polynomial-time deterministic algorithm with an asymptotic competitive ratio of nearly $1/n$ (Theorem~\ref{thm:det_lower}), which is the best possible.

We first observe an impossibility that the asymptotic competitive ratio is at most $1/n$ (Theorem~\ref{thm:upper_asymp_adverarial}).
Thus, our aim is to construct an asymptotically $1/n$-competitive algorithm.
If randomization is allowed, we can achieve it by simply allocating each item to an agent chosen uniformly at random.
We refer to this randomized algorithm as \Random.
Note that \Random{} is \emph{not} strictly $1/n$-competitive because the expected value of the minimum of random variables is \emph{not} equal to the minimum of the expected values of random variables.
We show that \Random{} guarantees $\Opt/n-O(\sqrt{\Opt\log\Opt})$ even for the adaptive-offline\footnotemark{} adversary, where $\Opt$ is the offline optimal value (Theorem~\ref{thm:rand_asymp_lower}).
\footnotetext{The adaptive-offline adversary chooses the next item based on the allocation chosen by the online algorithm thus far, and it obtains an offline optimal value for the resulting request items. }
Interestingly, this fact implies the existence of a \emph{deterministic} algorithm with the same guarantee~\cite{Ben-David1994}. However, the construction is not obvious. % due to the non-constructive proof. 
In fact, natural greedy algorithms are far from asymptotically $1/n$-competitive
(Theorem~\ref{thm:potential_upper} in Appendix).
Moreover, the natural round-robin procedure\footnotemark{} fails.
\footnotetext{In an offline setting, a round-robin procedure implies that agents take turns and choose their most preferred unallocated item. However, because we are dealing with online setting, we use this term to refer to a procedure in which the $j$th item is taken by agent $j\pmod{n}$.}
One disadvantage of these algorithms is that they output allocations that are too imbalanced and too balanced, respectively.
Moreover, it is unclear whether or not such a deterministic algorithm can be implemented to run in polynomial time.

We propose a novel derandomization method to obtain a polynomial-time deterministic algorithm with almost the same performance as \Random.
Our algorithm is based on the spirit of giving way to each other.
Upon the arrival of an item, our algorithm gives agents a chance to take it in ascending order with respect to the valuation of the item.
Each agent generously passes the chance in consideration of the agent's past assigned units.
Then, we can achieve the golden mean between allocations that are too balanced or too imbalanced, and we obtain the main result.
We believe that this technique is novel and will have further applications.
The advantage of our algorithm is that it does not require the information of the number of items nor an upper bound on the value of the items.
In addition, our analysis produces a consequence on another fairness notion called \emph{proportionality} (each of the $n$ agents receives a fraction at least $1/n$ of the entire items according to her valuation) in an asymptotic sense.

As an impossibility result,
we prove a stronger bound for deterministic algorithms: no deterministic online algorithm can attain $\Opt/n-\Omega((\Opt)^{\frac{1}{2}-\epsilon})$ for any $\epsilon>0$ where $\Opt$ is the offline optimal value (Theorem~\ref{thm:det_adversary}).
This bound implies that the performance of \Random{} is nearly optimal even when additive terms are taken into consideration.

We also show that
the strict competitive ratio of any deterministic algorithm is $0$ (Theorem~\ref{thm:det_upper}) and
the strict competitive ratio of the best randomized algorithm is $1/n^{\Theta(n)}$ (Theorems~\ref{thm:rand_strict_lower} and \ref{thm:oblivious_upper}).

\subsubsection{Unknown/known i.i.d.~arrival models}
% i.i.d.
Our main result for the i.i.d.\ arrival models is to provide an algorithm that outputs an asymptotically near-optimal allocation.
Our algorithm is the following simple one: upon the arrival of each item, allocate the item to the agent with the highest discounted value, where each agent's value of the item is exponentially discounted with respect to the total value received so far.
We prove that this algorithm with exponential base $(1-\epsilon/2)$ is $(1-\epsilon)$-competitive if the expected optimal value is larger than a certain value (Theorem~\ref{thm:iid_lower}).

We remark that our algorithm is based on a similar idea found in Devanur et al.~\cite{Devanur2019}, but this is not a naive application.
Devanur et al.~\cite{Devanur2019} provided an asymptotically $(1-\epsilon)$-competitive algorithm for a large class of resource allocation problems.
However, we have two difficulties when applying their algorithm to our problem.
One is that their algorithm requires the number $m$ of items to estimate the expected optimal value, but $m$ is unknown in our setting.
The other is that the setting of \cite{Devanur2019} deals with finite types of online items (i.e., each item is drawn from a discrete distribution) and their algorithm utilizes a linear programming (LP) solution;
by contrast, in our setting, there may exist infinite types of value vectors (i.e., a distribution can be continuous).
Our contribution is to resolve the above difficulties.
In fact, we do not use the LP in the algorithm (unlike the ones in \cite{Devanur2019}); we use it only in the analysis.
This makes our algorithm quite simple.
Note that our algorithm also does not require information about the total number of items nor an upper bound on the value of the items.

% strict 
For the strict competitive ratio, we show that even for the known i.i.d.~setting,
the strict competitive ratio of any algorithm must be exponentially small with respect to the number of agents (Theorem~\ref{thm:upper_known_iid}).

\medskip
The rest of this paper is organized as follows.
We formally define competitive ratios in Section~\ref{sec:preliminary-cr}.
We present our main algorithmic results for the adversarial and i.i.d.~arrival models in Sections~\ref{sec:adversarial-alg} and \ref{sec:i.i.d-algorithm}, respectively.
Then, in Sections~\ref{sec:hardness-adversarial} and \ref{sec:hardness-i.i.d.}, we present the impossibility results, which complement the algorithmic results.
We provide our concluding remarks in Section~\ref{sec:conclusion}.

\section{Preliminaries}
\label{sec:preliminary-cr}
To evaluate the performance of online algorithms, we use strict and asymptotic competitive ratios.
For an input sequence $\sigma$, let $\ALG(\sigma)$ and $\OPT(\sigma)$ respectively denote the egalitarian social welfares of the allocations obtained by an online algorithm $\ALG$ and an optimal offline algorithm $\OPT$ (here, $\ALG(\sigma)$ is a random variable if $\ALG$ is a randomized algorithm).
Then, the \emph{strict competitive ratio} and the \emph{asymptotic competitive ratio} for the adversarial arrival model are defined as
\begin{equation*}
  \inf_{\sigma}\frac{\mathbb{E}[\ALG(\sigma)]}{\OPT(\sigma)} \quad \text{ and } \quad
  \liminf_{\OPT(\sigma)\to\infty}\frac{\mathbb{E}[\ALG(\sigma)]}{\OPT(\sigma)},
\end{equation*}
respectively.
Here, the competitive ratios for randomized algorithms are defined by using an oblivious adversary.
The competitive ratios are at most $1$, and the larger values indicate better performance.
By the definition, the asymptotic competitive ratio of $\ALG$ is at least $\rho$ if $\mathbb{E}[\ALG(\sigma)]\ge \rho\cdot \OPT(\sigma)-o(\OPT(\sigma))$ for any input sequence $\sigma$.
Note that, in some literature (e.g.,~\cite{BE1998}),
the asymptotic competitive ratio of $\ALG$ is at least $\rho$ only when there is a constant $\alpha\ge 0$ such that $\mathbb{E}[\ALG(\sigma)]\ge \rho\cdot \OPT(\sigma)-\alpha$ for any input sequence $\sigma$.
We refer to this as the classical definition.

For the i.i.d.\ arrival model, we consider the distribution of input sequences $R(m,\cD)$ determined by a number of items $m$ and a distribution of value vectors $\cD$.
The \emph{strict competitive ratio} and the \emph{asymptotic competitive ratio} for the i.i.d.\ arrival model are similarly defined as
\begin{equation*}
  \inf_{m,\cD}\frac{\mathbb{E}_{\sigma\sim R(m,\cD)}[\ALG(\sigma)]}{\mathbb{E}_{\sigma\sim R(m,\cD)}[\OPT(\sigma)]}
  \quad
  \text{ and }
  \quad
  \liminf_{\OPT(\sigma)\to\infty}\frac{\mathbb{E}_{\sigma\sim R(m,\cD)}[\ALG(\sigma)]}{\mathbb{E}_{\sigma\sim R(m,\cD)}[\OPT(\sigma)]},
\end{equation*}
respectively.

\section{Algorithms for Adversarial Arrival}\label{sec:adversarial-alg}
In this section, we provide algorithms for the adversarial arrival model.
We first show a randomized algorithm that is asymptotically $1/n$-competitive in Section~\ref{subsec:adversarial-random} and then provide a deterministic algorithm with the same competitive ratio in Section~\ref{subsec:derandomize}.

\subsection{Randomized Algorithm}\label{subsec:adversarial-random}
A simple way to allocate items ``fairly'' is to allocate each item uniformly at random among all the agents.
We refer to this randomized algorithm as \Random{}.
One might think that it would be better to choose an agent who has a positive valuation for an item.
However, this does not perform better than \Random{} in the worst case scenario.
Furthermore, it turns out that \Random{} is a nearly optimal algorithm for the adversarial arrival model.

First, we prove that the asymptotic competitive ratio of \Random{} is at least $1/n$ by showing a slightly stronger statement.
\begin{theorem}\label{thm:rand_asymp_lower}
  For any adaptive adversary, \Random{} satisfies
  \begin{align}
    \mathbb{E}[\Random(\sigma)]\ge \frac{\Opt}{n}-O\left(\sqrt{\Opt\cdot\log\Opt}\right), \label{eq:rand_asymp_lower}
  \end{align}
  where $\sigma$ is the input sequence chosen by the adversary (depending on the stochastic behavior of \Random) and $\Opt=\mathbb{E}[\OPT(\sigma)]$.
\end{theorem}
\begin{proof}
  The adaptive adversary decides to request the next item or terminates depending on the sequence of allocation at each time so far.
  We use the symbol $e^\pi$ to denote the next item when the allocation sequence at the moment is $\pi$.
  Let $\Pi$ denote the set of all allocation sequences such that the adversary requests the next item.
  For each $\pi\in\Pi$, let $X_i^\pi$ be a random variable such that $X^\pi_i=1$ if \Random{} allocates $e^\pi$ to agent $i\in N$, and $X^\pi_i=0$ otherwise.
  In addition, let $Y^\pi$ be a random variable such that $Y^\pi=1$ if $e^\pi$ is requested (i.e., the allocation sequence chosen by \Random{} is $\pi$ at some moment), and $Y^\pi=0$ otherwise.
  As the allocation is totally uniformly at random, we have $\Pr[X_i^\pi=1 \, | \, Y^\pi=1]=1/n$ for all $i\in N$, and $\Pr[Y^\pi=1] = 1/n^{|\pi|}$, where $|\pi|$ denotes the length of $\pi$ (i.e., the number of items allocated so far).

  The total utility of agent $i$ is $S_i=\sum_{\pi\in\Pi}v_i(e^\pi) X_i^\pi Y^\pi$,
  and the expected utility of $i$ is $\mathbb{E}[S_i]=\sum_{\pi\in\Pi}\frac{v_i(e^\pi)}{n^{|\pi|+1}}$.
  Let $\mu_i=\mathbb{E}[S_i]$ for each $i\in N$, and let $\mu_{\min}=\min_{i\in N}\mu_i$.
  Then, the expected optimal value $\Opt$ is at most
  \begin{align}
    \Opt
     & \le \mathbb{E}\left[\min_{i\in N}\sum_{\pi\in\Pi}v_i(e^\pi) Y^\pi\right]
    \le \min_{i\in N}\mathbb{E}\left[\sum_{\pi\in\Pi}v_i(e^\pi) Y^\pi\right]    \\
     & \le \min_{i\in N}\sum_{\pi\in\Pi}\frac{v_i(e^\pi)}{n^{|\pi|}}
    = \min_{i\in N} n\cdot \mu_i = n\cdot \mu_{\min}.
  \end{align}

  We apply the Chernoff bound:
  since each $S_i$ ($i\in N$) satisfies $0\le S_i\le 1$, %and then 
  we have
  \begin{align}\label{eq:Chernoff}
    \Pr\left[S_i\le (1-\delta)\cdot\mu_i\right]\le \exp(-\mu_i \delta^2/2).
  \end{align}
  for all $\delta>0$.
  By setting $\delta=\sqrt{(2\log (n\mu_i))/\mu_i}$ in \eqref{eq:Chernoff}, we see that
  \begin{align}
    \Pr\bigl[S_i\le \mu_i-\sqrt{2\mu_i\log(n\mu_i)}\bigr]
     & =\Pr\bigl[S_i\le (1-\sqrt{(2\log (n\mu_i))/\mu_i})\cdot \mu_i\bigr]             \\
     & \le \exp\left(-\mu_i\cdot\frac{2\log(n\mu_i)/\mu_i}{2}\right)=\frac{1}{n\mu_i}.
  \end{align}
  Furthermore, by the union bound, the probability that $S_i\le \mu_i-\sqrt{2\mu_i\log(n\mu_i)}$ holds for some $i$ is at most $\sum_{i\in N}\frac{1}{n\mu_i}\le \frac{1}{\mu_{\min}}$.
  Without loss of generality, we may assume that $\mu_{\min}\ge 4n$ since we are analyzing asymptotic behavior.
  As $x-\sqrt{2x\log(nx)}$ is monotone increasing for $x\ge 4n$, we obtain
  \begin{align}
    \mathbb{E}\left[\min_{i\in N} S_i\right]
     & \ge \min_{i\in N} \bigl(\mu_i-\sqrt{2\mu_i\log(n\mu_i)}\bigr)\left(1-\frac{1}{\mu_{\min}}\right) \\
     & = \bigl(\mu_{\min}-\sqrt{2\mu_{\min}\log(n\mu_{\min})}\bigr)\left(1-\frac{1}{\mu_{\min}}\right)  \\
     & \ge \mu_{\min}-\sqrt{2\mu_{\min}\log(n\mu_{\min})}-1
    \ge \mu_{\min}-3\sqrt{\mu_{\min}\log(n\mu_{\min})}                                                  \\
     & \ge \frac{\Opt}{n}-3\sqrt{\frac{\Opt\cdot\log \Opt}{n}}
    =\frac{\Opt}{n}-O\left(\sqrt{\Opt\cdot\log \Opt}\right).
  \end{align}
\end{proof}

\begin{remark}
  In the classical definition of the asymptotic competitive ratio, \Random{} is at least $(1-\epsilon)/n$-competitive for any constant $\epsilon>0$ against adaptive-offline adversaries.
\end{remark}

We also analyze the strict competitive ratio of \Random.
For the strict competitive ratio, a deterministic algorithm can do almost nothing, but \Random{} attains $1/n^{O(n)}$ fraction of the optimal value.
Intuitively, this is because each agent obtains $\Omega(1/n)$ fraction of the value received in the optimal allocation with probability $\Omega(1/n)$. % (see Appendix~\ref{app:omitted} for the formal proof).
\begin{theorem}\label{thm:rand_strict_lower}
  The strict competitive ratio of \Random{} is at least $\frac{1}{n^{O(n)}}$ in the adversarial arrival model.
\end{theorem}
\begin{proof}
  Fix an input sequence $\sigma$,
  let $A^*$ be an offline optimum allocation for the instance and
  let $X_{ij}$ be the random variable that indicates that $e_j$ is allocated to $i$ by \Random{}.
  Note that $\OPT(\sigma)=\min_{i}v_i(A^*_i)$.
  For $i\in N$, consider the event $E_i$ such that \Random{} allocates $i$ at most $\frac{1}{2n}$ fraction of $A_i^*$ in terms of her valuation (i.e., $\sum_{e_j\in A_i^*}v_{i}(e_j)X_{ij}\le \frac{1}{2n}\cdot v_{i}(A_i^*)$).
  If none of $E_1,\dots,E_n$ occurs, then
  \begin{align}
    \Random(\sigma)
     & =\min_{i\in N}\sum_{e_j\in M}v_i(e)X_{ij}\ge \min_{i\in N}\sum_{e_j\in A_i^*}v_i(e)X_{ij} \\
     & >\min_{i\in N} \frac{v_i(A_i^*)}{2n}=\frac{1}{2n}\cdot\OPT(\sigma).
  \end{align}
  In addition, by Markov's inequality, we have
  \begin{align}
    \Pr\left[E_i\right]
     & = \Pr\left[1-\frac{\sum_{e_j\in A_i^*}v_{i}(e_j)X_{ij}}{v_i(A_i^*)}\ge 1-\frac{1}{2n}\right]
    \le \frac{1-\frac{1}{n}}{1-\frac{1}{2n}}=\frac{2n-2}{2n-1}.
  \end{align}
  As the events $E_1,\dots,E_n$ are independent, we obtain
  \begin{align}
    \mathbb{E}[\Random(\sigma)]
     & \ge \frac{\OPT(\sigma)}{2n}\cdot\prod_{i\in N}(1-\Pr[E_i])         \\
     & \ge \frac{\OPT(\sigma)}{2n}\cdot\left(1-\frac{2n-2}{2n-1}\right)^n
    =\frac{1}{n^{O(n)}}\cdot\OPT(\sigma).
  \end{align}
\end{proof}

\subsection{Derandomization}\label{subsec:derandomize}
It is well-known that there is no advantage to use randomization against adaptive-offline adversaries with respect to the competitive ratio~\cite{Ben-David1994}.
This implies the existence of a deterministic algorithm with the same guarantee as \Random.
However, the proof is not constructive, and hence it is not straightforward to obtain such a deterministic algorithm.
Moreover, there is no implication about running time.

A natural way to derandoimze \Random{} is a simple round-robin.
However, this fails due to the example in the Introduction (see Table~\ref{tbl:alternatively}).
Another approach is to estimate the optimal value, but this is impossible in the adversarial setting.
Moreover, we can prove that allocating new arriving item $e$ to the agent who maximize $\phi(v_i(A_i\cup\{e\}))-\phi(v_i(A_i))$ is not asymptotically $1/n$-competitive for any monotone increasing function $\phi$
(see Appendix~\ref{sec:greedy} for more details).

Our approach is to classify items into (infinitely many) types and aim to allocate almost the same number of items of each type to each agent.
Fixing a positive real $\epsilon$, we denote $\ind(x)=\lfloor\log_{1-\epsilon} x\rfloor$, where $\ind(0)=\infty$.
We define a \emph{type} of an item $e$ as a vector $(\ind(v_1(e)),\dots,\ind(v_n(e)))$.
Note that an agent with a smaller $\ind(x)$ has a higher valuation.
Now, our task is to schedule the order of allocation for each type of items.
If there are only $2$ agents, applying the round-robin procedure independently for each type (in which the first item is allocated to the agent who wants it more than the other) is asymptotically $(1-\epsilon)/2$-competitive.
However, in general, such a simple round-robin in a particular type
may result in a too unbalanced allocation as shown in Table~\ref{tbl:type_allocation3_bad}.
Thus, we introduce a sophisticated procedure to avoid such an unbalanced allocation.

\begin{table}[htb]
  \begin{minipage}{.55\textwidth}
    \centering
    \caption{Too unbalanced allocation $(n=3)$}\label{tbl:type_allocation3_bad}
    \tabcolsep=1mm
    \scalebox{1}{
      \begin{tabular}{c|ccccccc}
        \toprule
        $j$              & 1            & 2            & 3            & 4            & 5            & 6            & $\cdots$ \\\midrule
        $\ind(v_1(e_j))$ & \marked{$0$} & \marked{$0$} & \marked{$0$} & \marked{$0$} & \marked{$0$} & \marked{$0$} & $\cdots$ \\
        $\ind(v_2(e_j))$ & $1$          & $2$          & $1$          & $3$          & $1$          & $4$          & $\cdots$ \\
        $\ind(v_3(e_j))$ & $2$          & $1$          & $3$          & $1$          & $4$          & $1$          & $\cdots$ \\
        \bottomrule
      \end{tabular}
    }
  \end{minipage}%
  \begin{minipage}{.45\textwidth}
    \centering
    \caption{Our allocation $(n=3)$}\label{tbl:type_allocation3}
    \tabcolsep=1mm
    \scalebox{1}{
      \begin{tabular}{c|ccccccc}
        \toprule
        $j$              & 1            & 2            & 3            & 4            & 5            & 6            & $\cdots$ \\\midrule
        $\ind(v_1(e_j))$ & \marked{$0$} & \marked{$0$} & $0$          & $0$          & \marked{$0$} & \marked{$0$} & $\cdots$ \\
        $\ind(v_2(e_j))$ & $1$          & $2$          & \marked{$1$} & $3$          & $1$          & $4$          & $\cdots$ \\
        $\ind(v_3(e_j))$ & $2$          & $1$          & $3$          & \marked{$1$} & $4$          & $1$          & $\cdots$ \\
        \bottomrule
      \end{tabular}
    }
  \end{minipage}
\end{table}

We describe our novel technique of derandomization.
Suppose that the type of an arriving item $e$ is $(w_1,w_2,\dots,w_n)$ with $w_1\le w_2\le\dots\le w_n$.
By the definition of $\ind$, we have $(1-\epsilon)^{w_i+1}<v_i(e)\le (1-\epsilon)^{w_i}$.
Our algorithm gives agent $n$, who has the smallest value for $e$, a chance to receive $e$.
She obtains $e$ if she has passed previous $n-1$ chances to receive items of type $(w_1,\dots,w_n)$,
and passes the chance otherwise.
If agent $n$ passed the chance, then the algorithm gives agent $n-1$ a chance.
Agent $n-1$ obtains $e$ if she has passed previous $n-2$ chances to receive items of type $(w_1,\dots,w_{n-1},w_n')$ with some $w_n'\ge w_{n-1}$.
Note that $w'_n$ can vary.
For example, $n=3$ and if agent $2$ passes an item of type  $(w_1,w_2, \alpha)$, and agent $3$ passes a next item that has type $(w_1,w_2, \beta)$, then agent $2$ obtains the item.
Our algorithm repeats this procedure.
In general, if agents $n,n-1,\dots,i+1$ passed the chances, then the algorithm gives agent $i$ a chance to receive $e$.
Agent $i$ obtains $e$ if she has passed the previous $i-1$ chances to receive items of type $(w_1,\dots,w_{i},w_{i+1}',\dots,w_n')$ with some $w_i\le w_{i+1}'\le\dots\le w_{n}'$.
Note that the item $e$ is allocated to some agent. At least, agent $1$ obtains $e$ if she receives the chance.
See Table~\ref{tbl:type_allocation3} for an example of allocation by our algorithm.
We present a formal description in Algorithm~\ref{alg:det}.

\begin{algorithm}[htb]
  \caption{Asymptotically $(1-\epsilon)/n$-competitive deterministic algorithm (adversarial model)}\label{alg:det}
  Let $A_i \ot \emptyset$ for each $i\in N$\;
  \ForEach{\(e_j\ot e_1,e_2,\dots,e_m\)}{
  Let $\tau^j$ be a permutation over $N$ such that $v_{\tau(1)}(e_j)\ge v_{\tau(2)}(e_j)\ge\dots\ge v_{\tau(n)}(e_j)$\;
  Let $w^j_{i} \ot \ind(v_{\tau^j(i)}(e_j))$ for each $i\in N$\;
  \For{$i\ot n,n-1,\dots,1$}{
    Increment $x(\tau^j;w^j_1,w^j_2\dots,w^j_i)$ by $1$ (if the variable is undefined, then set it as $1$)\;
    \If{$x(\tau^j;w^j_1,w^j_2,\dots,w^j_i)=i$}{
      Allocate $e_j$ to agent $\tau^j(i)$ (i.e., $A_{\tau^j(i)}\ot A_{\tau^j(i)}\cup \{e_j\})$\;
      Set $x(\tau^j;w^j_1,w^j_2\dots,w^j_i)\ot 0$\;
      \Break\;
    }
  }
  }
\end{algorithm}

It is not difficult to see that Algorithm~\ref{alg:det} can be implemented to run in polynomial-time.
We prove the following statement.
\begin{theorem}\label{thm:det_lower}
  For any positive real $\epsilon<1$ and any input sequence $\sigma$, Algorithm~\ref{alg:det} returns an allocation $A$ such that $v_i(A_i)\ge \frac{1-\epsilon}{n}v_i(M)-\frac{(n!)^2}{\epsilon^{n}}$ for all $i\in N$ where $M$ is the set of items requested in $\sigma$.
\end{theorem}

This theorem implies that Algorithm~\ref{alg:det} is asymptotically $(1-\epsilon)/n$-competitive because
\begin{align}
  \min_{i\in N}v_i(A_i)
  \ge \min_{i\in N}\frac{1-\epsilon}{n}v_i(M)-\frac{(n!)^2}{\epsilon^n}
  \ge \frac{1-\epsilon}{n}\OPT(\sigma)-\frac{(n!)^2}{\epsilon^n}.
\end{align}
\begin{remark}
  %In addition, 
  The theorem also indicates that Algorithm~\ref{alg:det} finds a nearly proportional allocation, i.e., each agent receives at least nearly $1/n$-fraction of her valuation to the entire items.
\end{remark}

To prove the theorem, we show that the allocation is almost balanced regarding the number of items.
For a permutation $\tau$, index $k\in[n]$, and $\bm{w}=(w_1,w_2,\dots,w_k)\in\mathbb{Z}^k_{+}$ with $w_1\le w_2\le \dots\le w_k$, we denote $E^{\tau,k}(\bm{w})=\{e_j\mid \tau^j=\tau~\text{and}~(w^j_1,\dots,w^j_k)=\bm{w}\}$.
We remark that $\{E^{\tau, k}(\bm{w})\}_{\tau, \bm{w}}$ forms a partition of the entire item set for every $k\in[n]$.
\begin{lemma}\label{lem:det_upper}
  For any permutation $\tau$, index $k\in[n]$, and $\bm{w}=(w_1,\dots,w_k)\in\mathbb{Z}_+^k$ with $w_1\le w_2\le\dots\le w_k$, it holds that
  \[|A_{\tau(k)}\cap E^{\tau,k}(\bm{w})|\ge \frac{|E^{\tau,k}(\bm{w})|}{n}-1.\]
\end{lemma}
\begin{proof}
  We only discuss the chances regarding the items in $E^{\tau,k}(\bm{w})$.
  The number of chances that agent $\tau(n)$ receives is $|E^{\tau,k}(\bm{w})|$ because the algorithm gives the chance to $\tau(n)$ first for every item in $E^{\tau,k}(\bm{w})$.
  As $\tau(n)$ takes at most $1/n$ fraction of the chances, the number of chances that agent $\tau(n-1)$ receives is at least $(1-\frac{1}{n})|E^{\tau,k}(\bm{w})|=\frac{n-1}{n}|E^{\tau,k}(\bm{w})|$.
  Also, as $\tau(n-1)$ takes at most $1/(n-1)$ fraction of the chances, the number of chances that agent $\tau(n-2)$ receives is at least $(1-\frac{1}{n-1})\cdot \frac{n-1}{n}|E^{\tau,k}(\bm{w})|=\frac{n-2}{n}\cdot|E^{\tau,k}(\bm{w})|$ (if $k\le n-1$).
  Continuing the same argument, we can conclude that the number of chances that agent $\tau(i)$ receives is at least $\prod_{i'=i+1}^{n}(1-\frac{1}{i'})\cdot|E^{\tau,k}(\bm{w})|=\frac{i}{n}|E^{\tau,k}(\bm{w})|$ for every $i=n,n-1,\dots,k$ because whether she passes a chance or not is not affected by the items not in $E^{\tau,k}(\bm{w})$.
  As agent $\tau(k)$ receives an item if she has passed previous $k-1$ chances,
  we obtain
  $|A_{\tau(k)}\cap E^{\tau,k}(\bm{w})|\ge \left(\frac{k}{n}|E^{\tau,k}(\bm{w})|-(k-1)\right)\cdot\frac{1}{k}\ge \frac{1}{n}|E^{\tau,k}(\bm{w})|-1$.
\end{proof}

Now we are ready to prove Theorem~\ref{thm:det_lower}.
\begin{proof}[Proof of Theorem~\ref{thm:det_lower}]
  Let $i$ be an agent, $\tau$ be a permutation, and $k\in [n]$ be an index such that $i=\tau(k)$.
  Also, let $\bm{w}=(w_1,\dots,w_k)\in\mathbb{Z}_+^k$ with $w_1\le w_2\le\dots\le w_k$.
  Note that $(1-\epsilon)^{w_k+1}<v_i(e)\le (1-\epsilon)^{w_k}$ for every $e\in E^{\tau,k}(\bm{w})$.
  By Lemma~\ref{lem:det_upper}, we have
  \begin{align}
    v_i(A_i\cap E^{\tau,k}(\bm{w}))
     & \ge |A_i\cap E^{\tau,k}(\bm{w})|\cdot (1-\epsilon)^{w_k+1}                                 \\
     & \ge \left(\frac{1}{n}|E^{\tau,k}(\bm{w})|-1\right)\cdot (1-\epsilon)^{w_k+1}               \\
     & =   \frac{1-\epsilon}{n}\cdot |E^{\tau,k}(\bm{w})|(1-\epsilon)^{w_k} -(1-\epsilon)^{w_k+1} \\
     & \ge \frac{1-\epsilon}{n}v_i(E^{\tau,k}(\bm{w}))-(1-\epsilon)^{w_k+1}.\label{eq:det_lower}
  \end{align}
  By summing up \eqref{eq:det_lower} for all $\bm{w}'=(w'_1,\dots,w'_{k-1},w_k)$ with $w'_1\le\dots\le w'_{k-1}\le w_k$, we have
  \begin{align}
    \sum_{\bm{w}'}v_i(A_i\cap E^{\tau,k}(\bm{w}'))
     & \ge \sum_{\bm{w}'}\left(\frac{1-\epsilon}{n}v_i(E^{\tau,k}(\bm{w}'))-(1-\epsilon)^{w_k+1}\right)  \\
     & \ge \frac{1-\epsilon}{n}\sum_{\bm{w}'}v_i(E^{\tau,k}(\bm{w}'))-(w_k+1)^{k-1}(1-\epsilon)^{w_k+1}.
  \end{align}
  Finally, by summing up for all $\tau$, $k$ and $\bm{w}$, we obtain
  \begin{align}
    v_i(A_i)
     & = \sum_{\tau,k, \bm{w}}v_i(A_i\cap E^{\tau,k}(\bm{w}))                                                                                                          \\
     & \ge \frac{1-\epsilon}{n}\sum_{\tau,k,\bm{w}} v_i(E^{\tau,k}(\bm{w}))-\sum_{k\in N}\sum_{\tau:\,\tau(i)=k}\sum_{w_k=0}^{\infty}(w_k+1)^{k-1}(1-\epsilon)^{w_k+1} \\
     & \ge \frac{1-\epsilon}{n}v_i(M)-n!\cdot\sum_{\ell=0}^\infty(\ell+1)^{n-1}(1-\epsilon)^{\ell+1}                                                                   \\
     & \ge \frac{1-\epsilon}{n}v_i(M)-n!\cdot\sum_{\ell=0}^\infty(\ell+1)(\ell+2)\cdots(\ell+n-1)(1-\epsilon)^{\ell+1}                                                 \\
     & = \frac{1-\epsilon}{n}v_i(M)-\frac{n!\cdot (n-1)!}{\epsilon^n}\cdot(1-\epsilon)
    \ge \frac{1-\epsilon}{n}v_i(M)-\frac{(n!)^2}{\epsilon^n}.
  \end{align}
  Here, the second last equality holds because, for any $r$ with $|r|<1$,
  \begin{align}
    \frac{1}{(1-r)^n}
     & = (1+r+r^2+\cdots)^{n}
    = \sum_{\ell=0}^\infty\binom{\ell+n-1}{n-1}r^\ell                                       \\
     & = \frac{1}{(n-1)!}\cdot\sum_{\ell=0}^{\infty}(\ell+1)(\ell+2)\cdots(\ell+n-1)r^\ell.
  \end{align}
\end{proof}

\begin{remark}
  Algorithm~\ref{alg:det} works even if the upper bound of valuations is more than one and the algorithm does not know the upper bound.
  Let $\eta=\max_{i'\in N,~e\in M}v_{i'}(e)$.
  Note that $\ind(\eta)$ is a negative integer if $\eta>1$.
  Then, by summing up \eqref{eq:det_lower} for all types $\bm{w}\in\mathbb{Z}^n$ with $w_i\ge \ind(\eta)$ for all $i\in N$, we can obtain
  \begin{align}
    v_i(A_i)
     & = \sum_{\tau,k, \bm{w}}v_i(A_i\cap E^{\tau,k}(\bm{w}))                                                                                                                              \\
     & \ge \frac{1-\epsilon}{n}\sum_{\tau,k,\bm{w}} v_i(E^{\tau,k}(\bm{w}))-\sum_{k\in N}\sum_{\tau:\,\tau(i)=k}\sum_{w_k=\ind(\eta)}^{\infty}(w_k+1-\ind(\eta))^{k-1}(1-\epsilon)^{w_k+1} \\
     & \ge \frac{1-\epsilon}{n}v_i(M)-n!\cdot\sum_{\ell=0}^\infty(\ell+1)^{n-1}(1-\epsilon)^{\ell}\cdot(1-\epsilon)^{\ind(\eta)+1}
    \ge \frac{1-\epsilon}{n}v_i(M)-\frac{(n!)^2}{\epsilon^n}\cdot\eta
  \end{align}
  for each $i\in N$.
  Note that this bound is also useful for the case when $\eta\le 1$ because it implies a better guarantee.
\end{remark}

% doubling
\begin{remark}
  One may expect to design better performing algorithms by dynamically changing the value $\epsilon$ according to the current objective value.
  However, such a method does not work for the online max-min fair allocation problem. In fact, if an agent values $0$ for the items that come for a while at first, we essentially need to solve the problem for the other $n-1$ agents with a static $\epsilon$.
\end{remark}

\begin{remark}
  If $\OPT(\sigma)$ is known in advance (semi-online setting),
  Algorithm~\ref{alg:det} can output an allocation $A$ such that $\min_i v_i(A_i)\ge \frac{1}{n}\OPT(\sigma)-O((\OPT(\sigma))^{\frac{n}{n+1}})$ by setting $\epsilon=(\OPT(\sigma))^{1/(n+1)}$, i.e., Algorithm~\ref{alg:det} is asymptotically $1/n$-competitive in this setting.
\end{remark}

Finally, we discuss the difference between our results and the results of Benade et al~\cite{Benade2018}.
Recall that their deterministic algorithm outputs an allocation $A$ such that $v_i(A_i)\ge v_i(A_j)-O(\sqrt{m\log m})$ $(\forall i,j\in N)$.
This implies $v_i(A_i)\ge \frac{1}{n}v_i(M)-O(\sqrt{m\log m})$ for each $i\in N$, and hence $\min_{i\in N}v_i(A_i)\ge \frac{1}{n}\OPT(\sigma)-O(\sqrt{m\log m})$.
However, their algorithm has two drawbacks compared to ours.
One is that the additive term $O(\sqrt{m\log m})$ can be quite large compared to $\OPT(\sigma)$, for example, when the values of most items are almost zero for everyone.
The other one is that their algorithm requires fine-tuned parameters that depend on the number of items $m$ and an upper bound on the value of the items.
In contrast, our algorithm can be run independently of the number of items and the upper bound of the value of items, and our evaluation is independent of $m$. %the number of items.

\section{Algorithm for i.i.d.\ Arrival}\label{sec:i.i.d-algorithm}
In this section, we provide an algorithm for the i.i.d.\ arrival model, i.e., the value vector $\bm{v}$ of each item is drawn independently from a given distribution $\cD$.
We assume that the distribution $\cD$ and the total number $m$ of items are unknown to the algorithm.

For the strict competitive ratio, we can carry Theorem~\ref{thm:rand_strict_lower} for this case.
In what follows, we will analyze the asymptotic case.

One may expect that the round-robin procedure works well, but unfortunately it does not because, even if $n=2$ and $\cD$ is a distribution that takes $(1,\frac{1}{2})$ with probability $1$, the optimal value is $\frac{1}{3}m$ but the round-robin can achieve only $\frac{1}{4}m$.
We provide a simple algorithm that is asymptotically near-optimal.
Let $\epsilon >0$ be a fixed small constant.
When a new item $e_j$ arrives, our algorithm virtually discounts its value $v_i(e_j)$ for each agent $i$ by a factor $(1-\epsilon)^{v_i(A_i^{(j-1)})}$, where $A_i^{(j-1)}$ is the set of items allocated to $i$ so far.
Then, the algorithm allocates the item $e_j$ to the agent $i^{(j)}$ with the highest among discounted values, i.e., $i^{(j)}\in\argmax_{i} (1-\epsilon)^{v_i(A_i^{(j-1)})}v_i(e_j)$.
The discount factor leads to give a priority to an agent who has small utility at the moment.
We formally describe our algorithm in Algorithm~\ref{alg:iid}.
Note that the algorithm can be viewed as an application of the \emph{multiplicative weight update method}~\cite{Arora2005}, which is used to solve the experts problem.
However, the goals of the experts problem and the online max-min fair allocation problem are different, and no direct relationship can be found between them.
In addition, our algorithm does not use the information about the number of items, unlike the allocation algorithm given by Devanur et al.~\cite{Devanur2019}.

\begin{algorithm}[htb]
  \caption{Asymptotically $(1-2\epsilon)$-competitive deterministic algorithm (unknown i.i.d.~model)}\label{alg:iid}
  Let $A_i^{(0)}$ be the emptyset for each $i\in N$\;
  \ForEach{\(e_j\ot e_1,e_2,\dots,e_m\)}{
  Let $i^{(j)}$ be an agent in $\argmax_{i}\Bigl((1-\epsilon)^{v_i(A_i^{(j-1)})}\cdot v_{i}(e_j)\Bigr)$\;\label{line:iid_choice}
  Allocate $j$th item to $i^{(j)}$, i.e., $A^{(j)}_{i}\ot \begin{cases}A_{i}^{(j-1)}\cup\{e_j\}&(i=i^{(j)})\\ A^{(j-1)}_{i}&(i\ne i^{(j)})\end{cases}$ $(\forall i\in N)$\;
  }
\end{algorithm}

\begin{theorem}\label{thm:iid_lower}
  For any positive $\epsilon<1$, Algorithm~\ref{alg:iid} is $(1-2\epsilon)$-competitive if the expected optimal value is at least $\frac{2}{\epsilon^2}\log\frac{n}{\epsilon}$.
\end{theorem}
We prepare to prove Theorem~\ref{thm:iid_lower}.
We evaluate the performance of Algorithm~\ref{alg:iid} by using a linear programming problem that gives an upper bound of the optimal value.
For any realization of an input sequence $\sigma$, the optimal value $\OPT(\sigma)$ is equivalent to the optimal value of the following integer linear programming:
\begin{align}
  \begin{array}{rll}
    \max        & \lambda                                          &                                      \\
    \text{s.t.} & \lambda\le\sum_{j=1}^m v^\sigma_{ij}\cdot x_{ij} & (\forall i\in N),                    \\
                & \sum_{i\in N}x_{ij}=1                            & (\forall j\in [m]),                  \\
                & x_{ij}\in \{0,1\}                                & (\forall i\in N,\ \forall j\in [m]),
  \end{array}
  \tag{$\mathrm{IP}_\sigma$}\label{eq:IPR}
\end{align}
where we write $\bm{v}_j^\sigma=(v_{1j}^\sigma,\dots,v_{nj}^\sigma)$ to denote the value vector of the $j$th item in the instance $\sigma$.
The variable $x_{ij}$ corresponds a probability that agent $i$ receives the $j$th item.
Let $\Opt$ denote the expected optimal value $\mathbb{E}[\OPT(\sigma)]$.
Without loss of generality, we may assume that $\Opt>0$ because any algorithm is $1$-competitive if $\Opt=0$.

To analyze the performance of our algorithm, we consider an expected instance of the problem where everything happens as per expectation.
We construct a discretized problem because there are infinite possibilities of value vectors.
Let $\delta>0$ be a sufficiently small positive real.
For a real $u\in [0,1]$, define
\begin{align}
  \tau(u)=
  \begin{cases}
    \lceil\log_{1-\delta}u\rceil & \text{if }u\ge \delta\cdot\Opt/m, \\
    \infty                       & \text{if }u< \delta\cdot\Opt/m.
  \end{cases}
\end{align}
Note that $\tau(u)$ is in a finite set $\{0,1,\dots,\lceil\log_{1-\delta}(\delta\cdot\Opt/m)\rceil\}\cup\{\infty\}$, and we have
\begin{align}
  u\ge (1-\delta)^{\tau(u)}\ge (1-\delta)u \geq (1-\delta)u-\delta\cdot\Opt/m. \label{eq:type_ineq}
\end{align}
We define the \emph{type} of a value vector $\bm{u}=(u_1,\dots,u_n)\in [0,1]^n$, denoted by $\type(\bm{u})$, as $(\tau(u_1),\dots,\tau(u_n))$.\footnote{We note that the type defined here is slightly different from that defined in Section~\ref{subsec:derandomize}.}
Let $T$ be the set of all the possible types.
Then, the cardinality of $T$ is at most $(\lceil\log_{1-\delta}(\delta\Opt/m)\rceil+2)^n$, and hence it is finite.
The expected linear programming is defined as follows:
\begin{align}
  \begin{array}{rll}
    \max        & \lambda                                                                                                            &                                         \\
    \text{s.t.} & \lambda\le\sum_{\bm{t}\in T} m\cdot\Pr_{\bm{u}\sim\cD}[\type(\bm{u})=\bm{t}]\cdot(1-\delta)^{t_i}\cdot x_{i\bm{t}} & (\forall i\in N),                       \\
                & \sum_{i\in N}x_{i\bm{t}}=1                                                                                         & (\forall \bm{t}\in T),                  \\
                & x_{i\bm{t}}\ge 0                                                                                                   & (\forall i\in N,\ \forall \bm{t}\in T).
  \end{array}
  \tag{$\mathrm{LP}_\delta$}\label{eq:LPE}
\end{align}

We denote the optimal value of \eqref{eq:LPE} by $W_\delta$.
We prove that $W_\delta$ is not much smaller than $\Opt$.
\begin{lemma}\label{lem:delta_OPT}
  $W_\delta\ge (1-2\delta)\cdot \Opt$.
\end{lemma}
\begin{proof}
  For any $\sigma$, let $(\lambda^\sigma,(x_{ij}^\sigma)_{i\in N,\,j\in [m]})$ be an optimal solution for \eqref{eq:IPR}.
  For all $\sigma$, by the first constraint in \eqref{eq:IPR}, we have
  \begin{align}
    \lambda^\sigma\le \sum_{j=1}^m v_{ij}^\sigma\cdot x_{ij}^\sigma. \label{eq:lambda_ineq}
  \end{align}

  Define $\tilde{x}_{i\bm{t}} \coloneqq \mathbb{E}[x_{ij}^\sigma \,|\, \type(\bm{v}_j^\sigma)=\bm{t}]$,
  where the expectation is taken over the random choice of the instance $\sigma$, and the uniformly random choice of $j$ from $[m]$.
  As $\sum_{i\in N}x_{ij}^\sigma=1~(\forall j\in[m])$ and $x_{ij}^\sigma\in\{0,1\}~(\forall i\in N,\ \forall j\in[m])$, we have $\tilde{x}_{i\bm{t}}\ge 0~(\forall i\in N,\,\forall \bm{t}\in T)$ and $\sum_{i\in N}\tilde{x}_{i\bm{t}}=1~(\forall \bm{t}\in T)$.
  Moreover, for every $i\in N$, we have
  \begin{align}
    (1-2\delta)\Opt
     & = \mathbb{E}[(1-\delta)\lambda^\sigma]-\delta\Opt                                                                                                                                               \\
     & \le \mathbb{E}\left[\sum_{j=1}^m (1-\delta)v^\sigma_{ij}\cdot x_{ij}^\sigma\right]-\delta\Opt                                                          &  & (\text{by }\eqref{eq:lambda_ineq})  \\
     & \le \mathbb{E}\left[\sum_{j=1}^m \Bigl((1-\delta)v^\sigma_{ij}-\delta\Opt/m\Bigr)\cdot x_{ij}^\sigma\right]                                            &  & (\text{by }0\le x_{ij}^\sigma\le 1) \\
     & \le \mathbb{E}\left[\sum_{j=1}^m (1-\delta)^{\type(v_j^\sigma)_i}\cdot x_{ij}^\sigma\right]                                                            &  & (\text{by }\eqref{eq:type_ineq})    \\
     & = \sum_{\bm{t}\in T} m\cdot(1-\delta)^{t_i}\cdot \mathbb{E}[x^\sigma_{ij}\,|\, \type(\bm{v}^\sigma_{j})=\bm{t}]\cdot\Pr[\type(\bm{v}^\sigma_j)=\bm{t}]                                          \\
     & = \sum_{\bm{t}\in T} m\cdot\Pr[\type(\bm{v}^\sigma_j)=\bm{t}]\cdot(1-\delta)^{t_i}\cdot \tilde{x}_{i\bm{t}}                                                                                     \\
     & = \sum_{\bm{t}\in T} m\cdot\Pr\nolimits_{\bm{u}\sim \cD}[\type(\bm{u})=\bm{t}]\cdot(1-\delta)^{t_i}\cdot \tilde{x}_{i\bm{t}}.
  \end{align}
  Thus, $\bigl((1-2\delta)\Opt,(\tilde{x}_{i\bm{t}})_{i\in N,\,\bm{t}\in T}\bigr)$ is a feasible solution for \eqref{eq:LPE}.
  Hence, $W_\delta\ge (1-2\delta)\cdot\Opt$.
\end{proof}

Fix $\epsilon>0$. Let $X_{i,j}$ be random variables representing the values that agent $i$ obtains from the $j$th item in Algorithm~\ref{alg:iid}, i.e., $X_{i,j} = v_i(e_j)$ if agent $i$ receives the $j$th item, and $X_{i,j}=0$ otherwise.
The egalitarian social welfare of the allocation obtained by Algorithm~\ref{alg:iid} is $\min_{i\in N}\sum_{j=1}^m X_{i,j}$.
By the union bound and Markov's inequality, we have
\begin{align}
  \Pr\left[\min_{i\in N}\sum_{j=1}^m X_{i,j}\le (1-\epsilon)W_\delta\right]
   & \le \sum_{i\in N}\Pr\left[\sum_{j=1}^m X_{i,j}\le (1-\epsilon)W_\delta\right]                                                          \\
   & = \sum_{i\in N}\Pr\left[(1-\epsilon)^{\sum_{j=1}^m X_{i,j}}\ge (1-\epsilon)^{(1-\epsilon)W_\delta}\right]                              \\
   & \le \sum_{i\in N}\mathbb{E}\left[(1-\epsilon)^{\sum_{j=1}^m X_{i,j}}\right]/(1-\epsilon)^{(1-\epsilon)W_\delta}. \label{eq:iid_bound1}
\end{align}
In what follows, we prove that the rightmost value in \eqref{eq:iid_bound1} % above value 
is sufficiently small.
For $s=0,1,\dots,m$, let us define $\Phi(s)$ as follows:
\begin{align}
  \Phi(s)\coloneqq \sum_{i\in N}\mathbb{E}\left[(1-\epsilon)^{\sum_{j=1}^s X_{i,j}}\right]\cdot \Bigl(1-\frac{\epsilon W_\delta}{m}\Bigr)^{m-s}.
\end{align}
Note that the rightmost value in \eqref{eq:iid_bound1} is equal to $\Phi(m)/(1-\epsilon)^{(1-\epsilon)W_\delta}$.

\begin{lemma}\label{lem:potential_monotone}
  $\Phi(s)$ is monotone decreasing in $s$.
\end{lemma}
\begin{proof}
  Let $(W_\delta,(x^*_{i\bm{t}})_{i\in N,\,\bm{t}\in T})$ be an optimal solution of \eqref{eq:LPE}.
  To prove the lemma, we use the following algorithm as a baseline: allocating each item of type $\bm{t}$ to agent $i$ with probability $x^*_{i\bm{t}}$ regardless of previous allocations.
  Similarly to $X_{i,j}$, let $Y_{i,j}$ be random variables representing the values that agent $i$ obtains from the $j$th item in the baseline algorithm, i.e., $Y_{i,j} = v_i(e_j)$ if agent $i$ receives the $j$th item, and $Y_{i,j}=0$ otherwise.

  Then, for $s=0,1,\dots,m-1$, we have
  \begin{align}
    \Phi(s+1)
     & = \sum_{i\in N}\mathbb{E}\left[(1-\epsilon)^{\sum_{j=1}^{s+1} X_{i,j}}\right]\cdot \Bigl(1-\frac{\epsilon W_\delta}{m}\Bigr)^{m-s-1}                                              \\
     & = \sum_{i\in N}\mathbb{E}\left[(1-\epsilon)^{\sum_{j=1}^s X_{i,j}}\cdot (1-\epsilon)^{X_{i,s+1}}\right]\cdot \Bigl(1-\frac{\epsilon W_\delta}{m}\Bigr)^{m-s-1}                    \\
     & \le \sum_{i\in N}\mathbb{E}\left[(1-\epsilon)^{\sum_{j=1}^s X_{i,j}}\cdot (1-\epsilon X_{i,s+1})\right]\cdot \Bigl(1-\frac{\epsilon W_\delta}{m}\Bigr)^{m-s-1}                    \\
     & \le \sum_{i\in N}\mathbb{E}\left[(1-\epsilon)^{\sum_{j=1}^s X_{i,j}}\cdot (1-\epsilon Y_{i,s+1})\right]\cdot \Bigl(1-\frac{\epsilon W_\delta}{m}\Bigr)^{m-s-1}                    \\
     & = \sum_{i\in N}\mathbb{E}\left[(1-\epsilon)^{\sum_{j=1}^s X_{i,j}}\right]\cdot\mathbb{E}\left[1-\epsilon Y_{i,s+1}\right]\cdot \Bigl(1-\frac{\epsilon W_\delta}{m}\Bigr)^{m-s-1}  \\
     & \le \sum_{i\in N}\mathbb{E}\left[(1-\epsilon)^{\sum_{j=1}^s X_{i,j}}\right]\cdot\left(1-\epsilon \frac{W_\delta}{m}\right)\cdot \Bigl(1-\frac{\epsilon W_\delta}{m}\Bigr)^{m-s-1} \\
     & = \sum_{i\in N}\mathbb{E}\left[(1-\epsilon)^{\sum_{j=1}^s X_{i,j}}\right]\cdot\Bigl(1-\frac{\epsilon W_\delta}{m}\Bigr)^{m-s}
    = \Phi(s),
  \end{align}
  where the first inequality holds by $X_{i,s+1}\in [0,1]$ and $(1-\epsilon)^x\le 1-\epsilon x~(\forall x\in[0,1])$,
  the second inequality holds by the choice of line~\ref{line:iid_choice} in Algorithm~\ref{alg:iid},
  and the third inequality holds by
  \begin{align}
    \mathbb{E}[Y_{i,s+1}]
     & =   \mathbb{E}_{\bm{u}\sim \cD}[u_i\cdot x^*_{i\type(\bm{u})}]
    \ge \sum_{\bm{t}\in T}\Pr\nolimits_{\bm{u}\sim \cD}[\type(\bm{u})=\bm{t}]\cdot(1-\delta)^{t_i}\cdot x_{i\bm{t}}^*
    \ge W_\delta/m.
  \end{align}
\end{proof}

\begin{lemma}\label{lem:potential_zero}
  $\Phi(0)/(1-\epsilon)^{(1-\epsilon)W_\delta}\le n\cdot e^{-\frac{\epsilon^2}{2}W_\delta}$
\end{lemma}
\begin{proof}
  By simple calculations, we have
  \begin{align}
    \frac{\Phi(0)}{(1-\epsilon)^{(1-\epsilon)W_\delta}}
    =\sum_{i\in N}\frac{\Bigl(1-\frac{\epsilon W_\delta}{m}\Bigr)^{m}}{(1-\epsilon)^{(1-\epsilon)W_\delta}}
    \le n\cdot \frac{e^{-\epsilon W_\delta}}{(1-\epsilon)^{(1-\epsilon)W_\delta}}
    \le n\cdot e^{-\frac{\epsilon^2}{2}W_\delta}
  \end{align}
  where the first inequality holds by $1-x\le e^{-x}$ for any $x$ and the second inequality follows from the fact that $\frac{1}{(1-\epsilon)^{(1-\epsilon)}}\le e^{\epsilon-\epsilon^2/2}$ for any $\epsilon \in[0,1)$
\end{proof}

We are ready to prove Theorem~\ref{thm:iid_lower}.
\begin{proof}[Proof of Theorem~\ref{thm:iid_lower}]
  By applying Lemmas~\ref{lem:potential_monotone} and~\ref{lem:potential_zero} to \eqref{eq:iid_bound1}, we see that
  \begin{align}
    \Pr\left[\min_{i\in N}\sum_{j=1}^m X_{i,j}\le (1-\epsilon)W_\delta\right]
    \le \frac{\Phi(m)}{(1-\epsilon)^{(1-\epsilon)W_\delta}}
    \le \frac{\Phi(0)}{(1-\epsilon)^{(1-\epsilon)W_\delta}}
    \le n\cdot e^{-\frac{\epsilon^2}{2}W_\delta}.
  \end{align}
  Hence, we obtain
  \begin{align}
    \mathbb{E}\left[\min_{i\in N}\sum_{j=1}^m X_{i,j}\right]
     & \ge (1-\epsilon)\cdot (1-n\cdot e^{-\frac{\epsilon^2}{2}W_\delta})W_\delta                    \\
     & \ge (1-\epsilon)\cdot (1-n\cdot e^{-\frac{\epsilon^2}{2}(1-2\delta)\Opt})\cdot(1-2\delta)\Opt
  \end{align}
  by Lemma~\ref{lem:delta_OPT}.
  As the inequality holds for every positive $\delta$, we have
  \begin{align}
    \mathbb{E}\left[\min_{i\in N}\sum_{j=1}^m X_{i,j}\right]
     & \ge (1-\epsilon)\cdot (1-n\cdot e^{-\frac{\epsilon^2}{2}\Opt})\Opt
    \ge (1-\epsilon)\cdot (1-\epsilon)\Opt
    \ge (1-2\epsilon)\Opt,
  \end{align}
  where the second inequality follows from the assumption that $\Opt \geq \frac{2}{\epsilon^2}\log\frac{n}{\epsilon}$.
\end{proof}

\begin{remark}
  Algorithm~\ref{alg:iid} works even if the upper bound of valuations is more than one and the algorithm does not know the upper bound.
  Let $\eta$ be an upper bound of the value of the items, i.e., $\Pr_{\bm{u}\sim\cD}[\max_{i\in N}u_i\le \eta]=1$.
  Also, let $\hat{v}_i(e)=v_i(e)/\eta$ for each agent $i\in N$ and item $e\in M$.
  Then, by considering $\hat{\epsilon}$ such that $(1-\hat{\epsilon})=(1-\epsilon)^{\eta}$, Algorithm~\ref{alg:iid} can be interpreted as allocating each item $e_j$ to an agent in $\argmax_{i\in N} (1-\hat{\epsilon})^{\hat{v}_i(A_i^{(j-1)})}\cdot \hat{v}_i(e_j)$.
  Hence, we can conclude that Algorithm~\ref{alg:iid} is $(1-2\hat{\epsilon})$-competitive if the expected optimal value is at least $\eta\cdot\frac{2}{\hat{\epsilon}^2}\log\frac{n}{\hat{\epsilon}}$.
  Note that this bound is also useful for the case when $\eta\le 1$ because it implies a better guarantee.
\end{remark}

\begin{remark}
  The analysis in Theorem~\ref{thm:iid_lower} implies the following \emph{regret} bounds.
  If $\Opt$ is known in advance (semi-online setting),
  Algorithm~\ref{alg:iid} can attain $\mathbb{E}\bigl[\min_i \sum_{j=1}^m X_{i,j}\bigr]=\Opt-O(\sqrt{\Opt\log\Opt})$ by setting $\epsilon=2\sqrt{\frac{\log\Opt}{\Opt}}$.
  If the number of items $m$ is known in advance,
  Algorithm~\ref{alg:iid} can attain $\mathbb{E}\bigl[\min_i \sum_{j=1}^m X_{i,j}\bigr]=\Opt-O(\sqrt{m\log m})$ by setting $\epsilon=2\sqrt{\frac{\log m}{m}}$ because $\Opt\le m/n$.
\end{remark}

\section{Impossibilities for Adversarial Arrival}\label{sec:hardness-adversarial}
In this section, we provide upper bounds of competitive ratios for the adversarial arrival model.

\subsection{Asymptotic Competitive Ratio}
We show the upper bound of the asymptotic competitive ratio, which implies that the asymptotic competitive ratio of \Random{} is the best possible.
\begin{theorem}\label{thm:upper_asymp_adverarial}
  The asymptotic competitive ratio of any randomized algorithm is at most $1/n$ in the adversarial arrival model.
\end{theorem}
\begin{proof}
  We use Yao's principle.
  It suffices to prove that, for any $M>0$, there exists a probability distribution of input sequences $\sigma$ such that
  $\mathbb{E}[\OPT(\sigma)]>M$ and $\mathbb{E}[\ALG(\sigma)]\le\mathbb{E}[\OPT(\sigma)]/n$ for any deterministic algorithm $\ALG$.

  Let $k$ be an integer greater than $M$.
  Let us consider a distribution that chooses the following input sequence for each $i \in N$ with probability $1/n$:
  the first $k$ items have a value vector $\bm{1}$ and the subsequent $(n-1)k$ items have value vector $\bm{1}-\chi_i$, that is, agent $i$ has value $0$ and others have value $1$.
  Then we have $\mathbb{E}[\OPT(\sigma)]=k>M$ and $\mathbb{E}[\ALG(\sigma)]\le k/n=\mathbb{E}[\OPT(\sigma)]/n$.
\end{proof}

We show a stronger upper bound of the asymptotic competitive ratio for deterministic algorithms.
The upper bound implies that, for any positive reals $\epsilon~(<1/10)$, there exist input sequences $\sigma$ such that
$\ALG(\sigma)\le \frac{1}{n}\OPT(\sigma)-\Omega((\OPT(\sigma))^{\frac{1}{2}-\epsilon})$.
(This bound can be obtained by setting $\frac{1}{2}-\epsilon=\frac{1}{3-r}$.)
We construct an adversarial item sequence based on the idea inspired by Benade et al.~\cite{Benade2018}.
The sequence indicates that the competitive ratio of an algorithm will be far from $1/n$ if it tries to allocate in an overly balanced way.
\begin{theorem}\label{thm:det_adversary}
  Suppose that $n\ge 2$. Let $r$ be a real such that $1/2<r<1$.
  For any deterministic algorithm $\ALG$ and a positive real $c$,
  there exists an input sequence $\sigma$ such that $\ALG(\sigma)\le \OPT(\sigma)/n -c$ and $\OPT(\sigma)=O(c^{3-r})$.
\end{theorem}
\begin{proof}
  Fixing a deterministic algorithm $\ALG$ and a real $c>0$, we construct an adversary that gives the upper bound.
  Let $M^{(j)}$ be the set of the first $j$ items, and
  let $A_i^{(j)}$ be the allocation to agent $i\in N$ by the algorithm for $M^{(j)}$.
  The adversary keeps a deficiency state for each agent.
  The state is $0$ at the beginning for each agent.
  In each round, it increases by $1$, and afterwards, it decreases by $n$ if the agent receives an item in the round.
  The larger the deficiency state is, the less the value of the next item is.
  By definition, the sum of the deficiency states over the agents is $0$ in every round.
  Formally, let $s_i^{(j)}=(j-1)-n\cdot|A_i^{(j-1)}|$ be the state of $i\in N$ just before arriving $j$th item.
  Define
  \begin{align}
    \lambda(s)\coloneqq
    \begin{cases}
      (s+1)^r-s^r & (s\ge 0), \\
      1           & (s<0).
    \end{cases}
  \end{align}
  Then, the value of $j$th item is $\lambda(s_i^{(j)})$ for each $i\in N$ (see Table~\ref{tbl:det_adversary}).
  Roughly speaking, if an agent receives an item with value $\lambda(s)$, the agent has not received $(n-1)$ items with values $\lambda(s-1),\lambda(s-2),\dots,\lambda(s-(n-1))$.
  The adversary continues to request such items as long as $v_i(A_i^{(j)})\ge v_i(M^{(j)})/n-c$ for all $i$.
  If $v_{i^*}(A_{i^*}^{(j^*)})\le v_{i^*}(M^{(j^*)})/n-c$ for some $i^*\in N$ just after the $j^*$th item is allocated, then the adversary requests $(n-1)\cdot j^*$ items with value vector $\bm{1}-\chi_{i^*}$ and stop.
  Note that the optimal value is $\OPT(\sigma)=v_{i^*}(M_{i^*}^{(j^*)})$.
  The adversary is formally described in Algorithm~\ref{alg:adversary}.

  \begin{table}[ht]
    \centering
    \caption{An example of adversarial input ($n=3$)}\label{tbl:det_adversary}
    \tabcolsep=1mm
    \begin{tabular}{c|ccccccc}
      \toprule
      $j$        & 1                     & 2                     & 3                      & 4                     & 5                     & 6                     & $\cdots$ \\\midrule
      $v_1(e_j)$ & \marked{$\lambda(0)$} & $\lambda(-2)$         & \marked{$\lambda(-1)$} & $\lambda(-3)$         & $\lambda(-2)$         & $\lambda(-1)$         & $\cdots$ \\
      $v_2(e_j)$ & $\lambda(0)$          & \marked{$\lambda(1)$} & $\lambda(-1)$          & $\lambda(0)$          & $\lambda(1)$          & \marked{$\lambda(2)$} & $\cdots$ \\
      $v_3(e_j)$ & $\lambda(0)$          & $\lambda(1)$          & $\lambda(2)$           & \marked{$\lambda(3)$} & \marked{$\lambda(1)$} & $\lambda(-1)$         & $\cdots$ \\
      \bottomrule
    \end{tabular}
  \end{table}

  \begin{algorithm}[htb]
    \caption{Adversarial construction of an item sequence}\label{alg:adversary}
    Let $M^{(0)}\ot \emptyset$ and $A_i^{(0)}\ot \emptyset$ for each $i\in N$\;
    \For{\(j\ot 1,2,\dots\)}{%
    Let $s_i^{(j)}\ot (j-1)-n\cdot|A_i^{(j-1)}|$ for every $i\in N$\;
    Set $v_{i}(e_j)\ot \lambda(s_i^{(j)})$ for all $i\in N$ and request $e_j$ to $\ALG$\;
    Let $i^{(j)}\in N$ be the agent that $j$th item is allocated by $\ALG$\;
    Let $M^{(j)}\ot M^{(j-1)}\cup\{e_j\}$ and $A_{i}^{(j)}\ot \begin{cases}A_i^{(j-1)}\cup\{e_j\}&(i=i^{(j)}),\\A_i^{(j-1)}&(i\ne i^{(j)})\end{cases}$ $(\forall i\in N)$\;
    \If{$v_{i}(A_{i}^{(j)})\le v_{i}(M^{(j)})/n-c$ for some $i$}{
      $i^*\ot i$ and $j^*\ot j$\;
      \Break\label{line:break}
    }
    }
    \For{\(j\ot j^*+1,j^*+2,\dots,n\cdot j^*\)}{
      Set $v_{i}(e_j)\ot \begin{cases}1&(i\ne i^*),\\0&(i=i^*)\end{cases}$ for each $i\in N$ and request $e_j$ to $\ALG$\;
    }
  \end{algorithm}

  It suffices to prove that the adversary stops in $O(c^{3-r})$ steps because, if it stops, then
  \begin{align}
    \ALG(\sigma)
    \le v_{i^*}(A_{i^*}^{(j^*)})
    \le \frac{1}{n}\cdot v_{i^*}(M^{(j^*)})-c
    \le \frac{1}{n}\cdot\OPT(\sigma)-c.
  \end{align}

  We distinguish between two cases: $s_i^{(j)}>(nc)^{1/r}$ for some $i\in N$ and $j=O(c^{3-r})$, or not.

  Suppose that $s_i^{(j)}>(nc)^{1/r}$ for some $i\in N$ and $j=O(c^{3-r})$, i.e., the allocation $A^{j-1}$ is very deficient for $i$.
  We repeatedly pick a bundle of $n$ items from $M^{(j-1)}$ by the following procedure:
  pick arbitrarily $e_p\in A_i^{(j-1)}$ and items $e_{\mu(p,1)},\dots,e_{\mu(p,n-1)}\in M^{(j-1)}\setminus A_i^{(j-1)}$ which are respectively requested in states $s_i^{(p)}-1,s_i^{(p)}-2,\dots,s_i^{(p)}-(n-1)$.
  We can choose such items because each state increases at most by $1$ in every round (see Figure~\ref{fig:det_adversary}).
  Note that,
  by $v_i(e_p)=v_i(e_{\mu(p,q)})-(\lambda(s_i^{(p)}-1)-\lambda(s_i^{(p)}))$ and $v_i(e_{\mu(p,1)})\le v_i(e_{\mu(p,2)})\le\dots\le v_i(e_{\mu(p,n-1)})$, we have
  \begin{align}
    v_i(e_p)
     & = \frac{1}{n}\left(v_i(e_p)+\sum_{t=1}^{n-1}\bigl(v_i(e_{\mu(p,1)})-(\lambda(s_i^{(p)}-1)-\lambda(s_i^{(p)}))\bigr)\right)                         \\
     & \le \frac{1}{n}\left(v_i(e_p)+\sum_{t=1}^{n-1}\bigl(v_i(e_{\mu(p,t)})-(\lambda(s_i^{(p)}-1)-\lambda(s_i^{(p)}))\bigr)\right)                       \\
     & = \frac{1}{n}\cdot v_i(\{e_p,e_{\mu(p,1)},\dots,e_{\mu(p,n-1)}\})-\frac{n-1}{n}\bigl(\lambda(s_i^{(p)}-1)-\lambda(s_i^{(p)})\bigr) \label{eq:adv1} \\
     & \le \frac{1}{n}\cdot v_i(\{e_p,e_{\mu(p,1)},\dots,e_{\mu(p,n-1)}\}). \label{eq:adv2}
  \end{align}
  If we remove as many of such bundles as possible from $M^{(j-1)}$, the remaining items in $M^{(j-1)}$ are the ones that are requested as states $0,1,\dots,s^{(j)}_i-1$.
  Hence, by \eqref{eq:adv2}, we obtain
  \begin{align}
    v_i(A_i^{(j-1)})
     & \le \sum_{e_p\in A_i^{(j-1)}} \frac{1}{n}\cdot v_i(\{e_p,e_{\mu(p,1)},\dots,e_{\mu(p,n-1)}\}) \\
     & = \frac{1}{n}\cdot \left(v_i(M^{(j-1)})-\sum_{t=0}^{s_i^{(j)}-1} \lambda(t)\right)
    = \frac{1}{n}\cdot v_i(M^{(j-1)})-\frac{(s_i^{(j)})^r}{n}
    < \frac{1}{n}\cdot v_i(M^{(j-1)})-c.
  \end{align}

  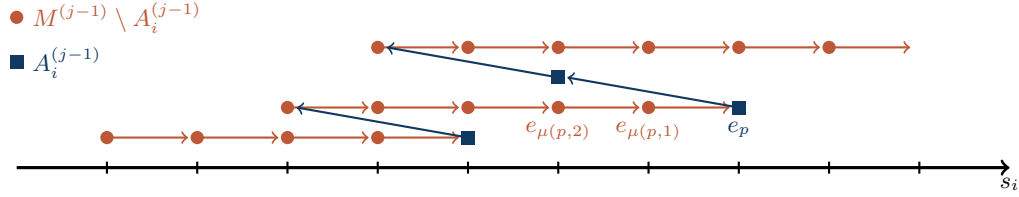
\begin{figure}[htb]
    \centering
    \begin{tikzpicture}[yscale=.4,xscale=1.2]
      \draw[->, very thick] (-1,0) -- (10,0) node[below] {$s_i$};
      \foreach \x in {0, 1, ..., 9}
        {
          \draw[thick] (\x,-0.2) -- (\x,0.2);
        }
      % forward
      \foreach \x/\y in {0/1,1/1,2/1,3/1,2/2,3/2,4/2,5/2,6/2,3/4,4/4,5/4,6/4,7/4,8/4}
        {
          \node[fill=myred,circle,minimum size=5pt,inner sep=0pt] at (\x,\y) {};
          \draw[->,thick,myred] (\x,\y) -- ($(\x,\y)+(.9,0)$);
        }
      % backward
      \foreach \x/\y in {4/1,7/2,5/3}
        {
          \draw node[fill=myblue,rectangle,minimum size=5pt,inner sep=0pt] at (\x,\y) {};
          \draw[->,thick,myblue] (\x,\y) -- ($(\x,\y)+(-1.9,1)$);
        }
      \node[fill=myblue,rectangle,minimum size=5pt,inner sep=0pt,label=right:\textcolor{myblue}{\small$A_i^{(j-1)}$}] at (-1,3.5) {};
      \node[fill=myred,circle,minimum size=5pt,inner sep=0pt,label=right:\textcolor{myred}{\small$M^{(j-1)}\setminus A_i^{(j-1)}$}] at (-1,5) {};
      \node[below=2pt,myblue] at (7,2) {$e_p$};
      \node[below=2pt,myred]  at (6,2) {$e_{\mu(p,1)}$};
      \node[below=2pt,myred]  at (5,2) {$e_{\mu(p,2)}$};
    \end{tikzpicture}
    \caption{An example of a sequence of changes for $s_i$ $(n=3)$}\label{fig:det_adversary}
  \end{figure}

  Next, suppose that $s_i^{(j)}\le (nc)^{1/r}$ for any $i\in N$ and $j=O(c^{3-r})$.
  Let us consider the same procedure as above for each $i\in N$ and $j=O(c^{3-r})$.
  If $s_i^{(j)}\ge 0$, we have
  \begin{align}
    v_i(A_i^{(j-1)})
     & \le \sum_{e_p\in A_i^{(j-1)}}\left(\frac{1}{n}\cdot v_i(\{e_p,e_{\mu(p,1)},\dots,e_{\mu(p,n-1)}\})-\frac{n-1}{n}\bigl(\lambda(s_i^{(p)}-1)-\lambda(s_i^{(p)})\bigr)\right) \\
    %& \le \frac{1}{n}v_i(M^{(j-1)})-\frac{1}{n}\sum_{t=0}^{s_i^{(j)}-1}\lambda(t)-\frac{n-1}{n}\sum_{e_p\in A_i^{(j-1)}}\bigl(\lambda(s_i^{(p)}-1)-\lambda(s_i^{(p)})\bigr) \\
     & \le \frac{1}{n}v_i(M^{(j-1)})-\frac{n-1}{n}\sum_{e_p\in A_i^{(j-1)}}\bigl(\lambda(s_i^{(p)}-1)-\lambda(s_i^{(p)})\bigr)                                                    \\
     & = \frac{1}{n}v_i(M^{(j-1)})-\frac{n-1}{n}\sum_{e_p\in A_i^{(j-1)}:\,s_i^{(p)}>0}\bigl(\lambda(s_i^{(p)}-1)-\lambda(s_i^{(p)})\bigr)                                        \\
     & \le \frac{1}{n}v_i(M^{(j-1)})-\frac{1}{2}\sum_{e_p\in A_i^{(j-1)}:\,s_i^{(p)}>0}\bigl(\lambda((nc)^{1/r}-1)-\lambda((nc)^{1/r})\bigr)                                      \\
     & \le \frac{1}{n}v_i(M^{(j-1)})-\frac{|\{e_p\in A_i^{(j-1)} \mid s_i^{(p)}>0\}|}{2}\bigl(\lambda((nc)^{1/r}-1)-\lambda((nc)^{1/r})\bigr)                                     \\
    %& \le \frac{1}{n}v_i(M^{(j-1)})-\frac{|\{e_{j'}\in M^{(j-1)} \mid s_i^{(j')}>0\}|}{2n}\bigl(\lambda((nc)^{1/r}-1)-\lambda((nc)^{1/r})\bigr)\\
     & \le \frac{1}{n}v_i(M^{(j-1)})-|\{e_p\in A_i^{(j-1)} \mid s_i^{(p)}>0\}|\cdot\frac{r(1-r)}{2c(nc+1)^{2-r}}\cdot c \label{eq:adv3}
    %& \le \frac{1}{n}v_i(M^{(j-1)})-\frac{|\{e_{j'}\in M^{(j-1)} \mid s_i^{(j')}>0\}|}{32(nc)^4}\cdot c,
  \end{align}
  where the first inequality holds by \eqref{eq:adv1} and the last inequality holds by
  \begin{align}
    \lambda(s-1)-\lambda(s)
     & \ge -\lambda'(s)=-r((s+1)^{r-1}-s^{r-1}) &  & (\because\text{$\lambda(s)$ is monotone convex}) \\
     & \ge r(1-r)(s+1)^{r-2}                    &  & (\because\text{$x^{r-1}$ is monotone convex}).
  \end{align}
  Let $h=\frac{2c(nc+1)^{2-r}}{r(1-r)}~(=\Theta(c^{3-r}))$.
  Now we are ready to show that $|\{e_p \in A_i^{(j-1)} \mid  s_i^{(p)}>0\}|\ge h$ and $s_i^{(j)}\ge 0$ for some $i$ and $j=O(c^{3-r})$. If this is the case, then \eqref{eq:adv3} implies $v_i(A_i^{(j-1)})\le \frac{1}{n}v_i(M^{(j-1)})-c$.

  Recall that $\sum_{i\in N}s_i^{(p)}=0$ for every $p$. Thus, at least one agent has a positive deficiency for every $p$ such that $p\not\equiv 1\pmod{n}$ because $s_i^{(p)}=0$ for all $i\in N$ if and only if all agents receive the same number of items at the beginning of $p$th round.
  Hence, $\sum_{i\in N}|\{p\in [j-1]\mid s_i^{(p)}>0\}|\ge (j-1)-\lceil (j-1)/n\rceil\ge (1-1/n)(j-1)-1\ge (j-3)/2$
  %$\sum_{i\in N}|\{e_{j'}\in M^{(j-1)}\mid s_i^{(j')}>0\}|\ge (j-1)-\lceil (j-1)/n\rceil\ge (1-1/n)j-1$
  and $|\{p\in [j-1]\mid s_i^{(p)}>0\}|\ge \frac{j-3}{2n}$ for some $i^*$.
  In addition, for each $i$,
  \begin{align}
    \max\{s_i^{(j)},\,0\}
    %&=\max\{(j-1)-n\cdot |A_i^{(j-1)}|,\,0\}\\
    %&\ge |\{p\in [j-1]\mid s_i^{(p)}>0\}|-\sum_{e_p\in A_i^{(j-1)}:\,s_i^{(p)}>0}\min\{n,\,s_i^{(p)}\}\\
     & \ge |\{p\in [j-1]\mid s_i^{(p)}>0\}|-n\cdot |\{e_{p}\in A_i^{(j-1)}\mid s_i^{(p)}>0\}|
  \end{align}
  because the total increments and decrements of the deficiency under the situation that the deficiency state is positive are $|\{p\in [j-1]\mid s_i^{(p)}>0\}|$ and  $n\cdot |\{e_{p}\in A_i^{(j-1)}\mid s_i^{(p)}>0\}|$, respectively.
  By the assumption that $s_{i}^{(j)}\le (nc)^{1/r}$, we have $|\{e_{p}\in A_i^{(j-1)}:\, s_{i^*}^{(p)}>0\}| \geq (\frac{j-3}{2n} - (nc)^{1/r})/n>h$ if $j\ge 2n(nh+(nc)^{1/r})+3$, which is $\Theta(c^{3-r})$ by $3-r>2>1/r$.
  %Let $i^*$ be such an $i$ and 
  Let $j^*=\min\bigl\{j\mid |\{e_{p}\in A_{i^*}^{(j-1)}:\, s_{i^*}^{(p)}>0\}|> h\bigr\}$.
  Then, $j^*=O(c^{3-r})$ and $s_{i^*}^{(j^*-1)}$ must be positive by the minimality of $j^*$, which implies the desired conclusion.
\end{proof}

\subsection{Strict Competitive Ratio}
We also discuss the strict competitive ratio.
First, the strict competitive ratio is $0$ for any deterministic algorithm.
\begin{theorem}\label{thm:det_upper}
  For any $n\ge 2$, the strict competitive ratio of any deterministic algorithm is $0$ in the adversarial arrival model.
\end{theorem}
\begin{proof}
  Fix an algorithm $\ALG$ and suppose that the value vector of the first item is $\bm{1}$.
  Without loss of generality, we may assume that the algorithm allocates the first item to the agent $1$.
  Suppose that the following $n-1$ items have value vector $(1,\dots,1,0)~(=\bm{1}-\chi_n)$.
  Then, the egalitarian social welfare of the optimal offline algorithm and the algorithm $\ALG$ are $1$ and $0$, respectively (see Table~\ref{tbl:det_upper}).
\end{proof}

\begin{table}[t]
  \begin{minipage}{.45\textwidth}
    \centering
    \caption{An instance for deterministic algorithm}\label{tbl:det_upper}
    \begin{tabular}{c|ccccc}
      \toprule
      $j$            & $1$          & $2$      & $\cdots$ & $n-1$    & $n$      \\ \midrule
      $v_{1}(e_j)$   & \marked{$1$} & $1$      & $\cdots$ & $1$      & $1$      \\
      $v_{2}(e_j)$   & $1$          & $1$      & $\cdots$ & $1$      & $1$      \\
      $\vdots$       & $\vdots$     & $\vdots$ &          & $\vdots$ & $\vdots$ \\
      $v_{n-1}(e_j)$ & $1$          & $1$      & $\cdots$ & $1$      & $1$      \\
      $v_{n}(e_j)$   & $1$          & $0$      & $\cdots$ & $0$      & $0$      \\
      \bottomrule
    \end{tabular}
  \end{minipage}%
  \quad
  \begin{minipage}{.45\textwidth}
    \centering
    \caption{An instance for randomized algorithm}\label{tbl:oblivious_upper}
    \begin{tabular}{c|ccccc}
      \toprule
      $j$                  & 1        & 2        & $\cdots$ & $n-1$    & $n$      \\ \midrule
      $v_{\tau(1)}(e_j)$   & $1$      & $0$      & $\cdots$ & $0$      & $0$      \\
      $v_{\tau(2)}(e_j)$   & $1$      & $1$      & $\cdots$ & $0$      & $0$      \\
      $\vdots$             & $\vdots$ & $\vdots$ &          & $\vdots$ & $\vdots$ \\
      $v_{\tau(n-1)}(e_j)$ & $1$      & $1$      & $\cdots$ & $1$      & $0$      \\
      $v_{\tau(n)}(e_j)$   & $1$      & $1$      & $\cdots$ & $1$      & $1$      \\
      \bottomrule
    \end{tabular}
  \end{minipage}%
\end{table}

Next, we observe that, for any randomized algorithm, the strict competitive ratio could be positive but at most $1/n!~(=1/n^{\Theta(n)})$.
This upper bound means that the strict competitive ratio of \Random{} is (almost) tight.
The proof is based on Yao's principle.
\begin{theorem}\label{thm:oblivious_upper}
  The strict competitive ratio of any randomized algorithm is at most $1/n!$ in the adversarial arrival model.
\end{theorem}
\begin{proof}
  It is sufficient to prove that there exists a probability distribution of input sequences $\sigma$ such that
  $\mathbb{E}[\OPT(\sigma)]=1$ and $\mathbb{E}[\ALG(\sigma)]\le \frac{1}{n!}$ for any randomized algorithm $\ALG$.
  Let $\tau$ be a permutation chosen uniformly at random from the set of permutations over $[n]$.
  Suppose that the input sequence consists of $n$ items $e_1,\dots, e_n$ with
  \begin{align}
    v_{i}(e_j)=
    \begin{cases}
      0 & \text{if }i\in\{\tau(1),\dots,\tau(j-1)\}, \\
      1 & \text{if }i\in\{\tau(j),\dots,\tau(n)\}
    \end{cases}
    \quad (\forall j\in [n]).
  \end{align}
  The egalitarian social welfare is $1$ only when $j$th item is allocated to agent $\tau(j)$ for all $j$ (and otherwise it is $0$).
  As an algorithm allocates $j$th item based only on the information about $\tau(1),\dots,\tau(j-1)$, it can allocate the item to $\tau(j)$ with probability at most $1/(n-j+1)$.
  Hence, the expected egalitarian social welfare that any algorithm can obtain is at most $1/n!$, despite the optimal egalitarian social welfare is always $1$.
\end{proof}

\section{Impossibilities for i.i.d.~Arrival}\label{sec:hardness-i.i.d.}
In this section, we provide an upper bound of competitive ratios for the i.i.d.\ arrival model.
As we show an asymptotic $(1-O(\epsilon))$-competitive algorithm in Section~\ref{sec:i.i.d-algorithm} for the unknown case, we only need to discuss the strict competitive ratio.
The following theorem states that the strict competitive ratio is at most $\frac{1}{e^{\Omega(n)}}$ even if the algorithm knows the distribution of value vectors and the number of items.
\begin{theorem}\label{thm:upper_known_iid}
  There exist a distribution over value vectors and a number of items such that the strict competitive ratio is at most $\frac{1}{e^{\Omega(n)}}$ for any algorithm in the i.i.d.\ arrival model.
\end{theorem}
\begin{proof}
  Without loss of generality, we may assume that $n$ is an even number.
  Suppose that the number of items $m$ equals $n$, and the value vector of an item is independently drawn from the following distribution:
  \begin{itemize}
    \item $\chi_i$ with probability $\frac{2}{3n}$ for each $i\in [n]$,
    \item $\psi_k\coloneqq \chi_{2k-1}+\chi_{2k}$ with probability $\frac{2}{3n}$ for each $k\in[n/2]$.
  \end{itemize}
  Note that each realization of the input sequence can be interpreted as a bipartite graph between agents and items where there exists an edge between agent $i$ and item $e_j$ if and only if $v_{i}(e_j)=1$.
  Under the above interpretation, the egalitarian social welfare of an allocation is $1$ if it corresponds to a perfect matching and $0$ otherwise.

  We bound the conditional probability that the egalitarian social welfare of an algorithm is $1$ under the condition that the optimal value is $1$.
  Suppose that the optimal value is $1$.
  Then, for each $k\in [n/2]$, exactly two of the value vectors of the input items must be in $\{\chi_{2k-1},\chi_{2k},\psi_k\}$ because these three value vectors are the only ones that are valuable to agent $2k-1$ or $2k$.
  Extracting such items in order of arrival, we see that the following seven patterns of inputs occur with equal probabilities:
  $(\chi_{2k-1},\chi_{2k})$, $(\chi_{2k},\chi_{2k-1})$, $(\chi_{2k-1},\psi_{k})$, $(\psi_{k},\chi_{2k-1})$, $(\chi_{2k},\psi_{k})$, $(\psi_{k},\chi_{2k})$, $(\psi_{k},\psi_{k})$.
  Note that if the realization is $(\psi_k,\chi_{2k-1})$ or $(\psi_k,\chi_{2k})$, the algorithm fails with probability at least $1/2$.
  Thus, the probability that the algorithm successfully allocates items to agents $2k$ and $2k-1$ is at most $6/7$.
  Considering for all $k\in [n/2]$, the algorithm successfully allocates all the items with probability at most $(6/7)^{n/2}=1/e^{\Omega(n)}$.
  Hence, the strict competitive ratio is at most $1/e^{\Omega(n)}$ for any algorithm.
\end{proof}

\section{Concluding remarks}\label{sec:conclusion}
% our results
In this paper, we have revealed asymptotic and strict competitive ratios of the online max-min fair allocation problem for the adversarial and i.i.d~arrival models.
Specifically, we designed polynomial-time deterministic algorithms that achieve asymptotically $\frac{1-\epsilon}{n}$-competitive for the adversarial arrival model and $(1-\epsilon)$-competitive for the i.i.d.\ arrival model, respectively, for any $\epsilon>0$.

We would like to mention a partial information model of our problem.
We have focused on the case where the values of agents for the current item are revealed before allocation.
The model where the values are revealed \emph{after} allocation (like the expert problem or the multi-armed bandit problem) also seems reasonable, but such a model is too restrictive.
In fact, by considering a distribution that takes $\chi_i$ with probability $1/n$ for each $i$, we can see that the asymptotic competitive ratio is at most $1/n$ even for the i.i.d.\ arrival model.
Even worse, by considering an adversarial arrival where the value vector of every item allocated to agent $i$ turns out to be $\bm{1}-(1-\epsilon)\chi_i$ with $\epsilon>0$, we can see that the asymptotic competitive ratio of any deterministic algorithm is at most $\epsilon$.

% future work
Finally, we discuss some possible future directions.
Although our algorithms are nearly optimal in an asymptotic sense, there is still room for improvement in additive terms.
Another important challenge is to design algorithms for the random order model.
Unlike the i.i.d.\ arrival model, it seems hard to construct an asymptotically $(1-\epsilon)$-competitive algorithm for the random order model because it inherently requires solving hard instances of the offline max-min fair allocation (the Santa Claus) problem.
Furthermore, constructing algorithms with best-possible guarantees simultaneously for the adversarial arrival and the i.i.d.\ arrival (best of both worlds) would be a crucial challenge.

\section*{Acknowledgments}
This research is part of the results of Value Exchange Engineering, a joint research project between Mercari, Inc. and the RIISE.
The first author was supported by JSPS KAKENHI Grant Number 20K19739 and JST PRESTO Grant Number JPMJPR2122.
The second author was supported by JSPS KAKENHI Grant Numbers 17K12646 and 21K17708.

%% Reference
\bibliography{fair}

\clearpage

\appendix

\section{Impossibility of Greedy-type Algorithms}\label{sec:greedy}
We may consider several variants of greedy algorithms, e.g., allocate the next item to maximize a power mean or the Nash product (geometric mean).
In general, greedy-type algorithms can be formalized
as algorithms that allocate the next item $e$ to agent $i$ maximizing $\phi(v_i(A_{i}\cup \{e\}))-\phi(v_i(A_i))$
where $\phi\colon\mathbb{R}_+\to\mathbb{R}_+$ is a monotone increasing function and $A_{i}$ is the set of items which agent $i$ receives so far.
One may expect that an almost optimum deterministic algorithm can be obtained by choosing an appropriate function $\phi$.
Such a method is known to work well in other problem settings, and indeed Benade et al.~\cite{Benade2018} successfully designed a deterministic online allocation algorithm such that the maximum envy is sublinear with respect to the number of items.
However, any greedy-type algorithm cannot achieve asymptotically $1/2$-competitive for the adversarial model with $n=2$ (despite the existence of asymptotically $1/2$-competitive deterministic algorithm).

\begin{theorem}\label{thm:potential_upper}
  The asymptotic competitive ratio of any greedy-type algorithm is at most $\frac{3-\sqrt{5}}{2}\approx 0.3820$ for the adversarial model with $n=2$.
\end{theorem}
\begin{proof}
  Fix a greedy-type algorithm.
  Let $\epsilon$ be the inverse of a sufficiently large positive integer.
  Suppose that the first $1/\epsilon^2$ items have value vector $(1,\epsilon)$.

  If the algorithm allocates at most $\frac{3-\sqrt{5}}{2\epsilon^2}$ of them to agent $2$, then consider the input sequence where the subsequent $1/\epsilon$ items have value vector $(1,0)$.
  The optimal solution for the sequence is that the first $1/\epsilon^2$ items are allocated to agent $2$ and the rest items are allocated to agent $1$.
  Thus, the optimal value is $1/\epsilon$. On the other hand, the egalitarian social welfare by the algorithm is at most $\frac{3-\sqrt{5}}{2\epsilon}$.
  Hence, the asymptotic competitive ratio is at most $\frac{3-\sqrt{5}}{2}$.\footnote{We remark that the optimal value in this case is $\Opt=1/\epsilon=\Theta(\sqrt{m})$, where $m$ is the number of items given so far. Thus, this fact does not preclude the existence of a greedy-type algorithm that guarantees $\frac{1}{n}\Opt-O(\sqrt{m})$.}

  Conversely, if the algorithm allocates at least $\frac{3-\sqrt{5}}{2\epsilon^2}$ of them to agent $2$ (i.e., at most $\frac{\sqrt{5}-1}{2\epsilon^2}$ of them to agent $1$),
  then consider the input sequence where the subsequent $\lfloor\frac{\sqrt{5}-1}{2\epsilon^2}\rfloor$ items have value vector $(1,1)$ and the following last $\lceil\frac{1+\sqrt{5}}{2\epsilon^2}\rceil$ items have value vector $(0,1)$.
  As the greedy-type algorithm allocates the items with value vector $(1,1)$ in a balanced manner, the egalitarian social welfare of the outcome is at most $\frac{\sqrt{5}-1}{2\epsilon^2}+\frac{1}{2\epsilon}+1$.
  Meanwhile, the optimal value is at least $\frac{1+\sqrt{5}}{2\epsilon^2}-1$ (which can be attained by allocating the first $\frac{1}{\epsilon^2}+\lfloor\frac{\sqrt{5}-1}{2\epsilon^2}\rfloor$ items to agent $1$ and the rest items to agent $2$).
  Hence, the asymptotic competitive ratio is at most
  \begin{align}
    \frac{\frac{\sqrt{5}-1}{2\epsilon^2}+\frac{1}{2\epsilon}+1}{\frac{1+\sqrt{5}}{2\epsilon^2}-1}
    \xrightarrow{\epsilon\to 0}
    \frac{\frac{\sqrt{5}-1}{2}}{\frac{1+\sqrt{5}}{2}}= \frac{3-\sqrt{5}}{2}.
  \end{align}
\end{proof}

In a similar way, we can prove that the asymptotic competitive ratio of any greedy-type algorithm is at most $\frac{2}{n+1+\sqrt{n^2+2n-3}}~(<1/n)$ for the adversarial model with $n$ agents.

\end{document}